\documentclass[conference,letterpaper]{IEEEtran}

\pdfoutput=1
\usepackage[cmex10]{amsmath} 
\usepackage{cite,amsfonts,amssymb,psfrag,paralist}
\usepackage{cancel}
\usepackage{graphicx}
\usepackage{epstopdf}
\usepackage{rotating}
\usepackage{dsfont}
\usepackage{color}
\usepackage{tikz}
\usetikzlibrary{plotmarks}
\usepackage{pgfplots}
\usetikzlibrary{calc}
\usetikzlibrary{shapes,arrows}
\usetikzlibrary{decorations.markings}
\usetikzlibrary{positioning}
\pgfplotsset{compat=1.10}
\usetikzlibrary{calc}
\usetikzlibrary{shapes,arrows}
\usetikzlibrary{decorations.markings}

\usepackage{bbm}
\usepackage{balance}

\usepackage{url,hyperref}
\hypersetup{colorlinks, citecolor=black, filecolor=black, linkcolor=black, urlcolor=black}

\usepackage[utf8]{inputenc} 
\usepackage[T1]{fontenc}
\usepackage{url}
\usepackage{ifthen}

\DeclareFontFamily{U}{mathx}{\hyphenchar\font45}
\DeclareFontShape{U}{mathx}{m}{n}{<-> mathx10}{}
\DeclareSymbolFont{mathx}{U}{mathx}{m}{n}
\DeclareMathAccent{\widebar}{0}{mathx}{"73}

\def\BibTeX{{\rm B\kern-.05em{\sc i\kern-.025em b}\kern-.08em T\kern-.1667em\lower.7ex\hbox{E}\kern-.125emX}}

\DeclareFontFamily{U}{mathx}{\hyphenchar\font45}
\DeclareFontShape{U}{mathx}{m}{n}{<-> mathx10}{}
\DeclareSymbolFont{mathx}{U}{mathx}{m}{n}

\newcommand{\Enc}{\mathsf{Enc}}
\newcommand{\Dec}{\mathsf{Dec}}
\newcommand{\Est}{\mathsf{Est}}

\newcommand{\overbar}[1]{\mkern 1.5mu\overline{\mkern-1.5mu#1\mkern-1.5mu}\mkern 1.5mu}

\usepackage{scalerel}
\usepackage{stackengine}
\stackMath
\def\hatgap{2pt}
\def\subdown{-2pt}
\newcommand\reallywidehat[2][]{%
  \renewcommand\stackalignment{l}%
  \stackon[\hatgap]{#2}{%
    \stretchto{%
      \scalerel*[\widthof{$#2$}]{\kern-.6pt\bigwedge\kern-.6pt}%
      {\rule[-5\textheight]{0.1ex}{\textheight}}
    }{0.5ex}
    _{\smash{\belowbaseline[\subdown]{\scriptstyle#1}}}%
  }}

\newtheorem{theorem}{Theorem}

\newtheorem{remark}{Remark}
\newtheorem{definition}{Definition}

\newtheorem{lemma}{Lemma}

\begin{document}
\title{Transmitter Actions for Secure Integrated Sensing and Communication}
\IEEEoverridecommandlockouts
\renewcommand\footnotemark{}
\renewcommand\footnoterule{}

\author{
  \IEEEauthorblockN{Truman Welling\textsuperscript{1}, Onur G\"unl\"u\textsuperscript{2}, and Aylin Yener\textsuperscript{1}}
    \IEEEauthorblockA{\textsuperscript{1}%
        Department of Electrical and Computer Engineering, The Ohio State University
    }
    \IEEEauthorblockA{\textsuperscript{2}%
        Information Theory and Security Laboratory (ITSL), Link{\"o}ping University\\ welling.78@buckeyemail.osu.edu, onur.gunlu@liu.se, yener@ece.osu.edu
    }

}

\maketitle

\begin{abstract}
     This work models a secure integrated sensing and communication (ISAC) system as a wiretap channel with action-dependent channel states and channel output feedback, e.g., obtained through reflections. The transmitted message is split into a common and a secure message, both of which must be reliably recovered at the legitimate receiver, while the secure message needs to be kept secret from the eavesdropper. The transmitter actions, such as beamforming vector design, affect the corresponding state at each channel use. The action sequence is modeled to depend on both the transmitted message and channel output feedback. For perfect channel output feedback, the secrecy-distortion regions are provided for physically-degraded and reversely-physically-degraded secure ISAC channels with transmitter actions. The corresponding rate regions when the entire message should be kept secret are also provided. The results are illustrated through characterizing the secrecy-distortion region of a binary example.
\end{abstract}

\section{Introduction} \label{sec:intro}
Modern communication systems have continuously evolved, improving on existing features and seeking ways to expand functionality. One promising facet currently being explored for future systems is joint communication and sensing \cite{NokiaGuysJCASTutorial,MariMichelleGJournal}. The integration of sensing and communication improves spectral and energy efficiencies of systems and reduces hardware cost \cite{liu2022isac}. The inherent nature of integrated sensing and communication (ISAC) is that sensing any target makes the communication signal available to them. Thus, security, such as in the context of information confidentiality\cite{tekin2005secureGMAWTC}, is a fundamental aspect for designing ISAC systems \cite{ourbinaryAWGNISAC,JCASwithSecurityTutorial, gunlu2023secureISAC}.

Secure ISAC systems have recently been considered in the literature. In \cite{ren2023beamSecureISAC}, a system securely communicating with a single user while sensing other targets uses artificial noise to obfuscate the message. Another approach in \cite{bazzi2024secureFullDuplex} uses artificial noise to secure a full duplex ISAC system. The work in \cite{gunlu2023secureISAC} models an ISAC system as a state-dependent wiretap channel with channel output feedback, where one user is an eavesdropper from whom some or all of the message should be kept secret. Sensing is performed at the transmitter by estimating the state based on the channel output feedback.

In this paper, we extend the secure ISAC model in \cite{gunlu2023secureISAC} to the action-dependent set up by introducing transmitter actions which affect the channel states, as in \cite{weissman2010actionCapacity}. The transmitter actions can define, e.g., the design of beamforming vectors with aim to improve the advantage of the legitimate receiver over the eavesdropper. Considering transmitter actions introduces a dependence between the states and the message, changing the secrecy-distortion regions and requiring more care when considering the dependence of the state sequence with other random variables in the proofs compared to \cite{gunlu2023secureISAC}.

The proposed secure ISAC model can be viewed as extensions of the wiretap channel with feedback models~\cite{AhlswedeCaiWTCwithFeedback,AsafCohenWTCwithFeedback,OurJSAITTutorial,HanVinckWTCwithFeedback,he-yener-fbsecrecy,GermanWTCwithGeneralizedFeedback,YHKimWTCwithFeedback,Tahmasbi2018} as well as channel feedback with actions \cite{zhang2019actionDependentfeedback, weissman2010actionCapacity}. One main difference, among others, between our model and those in \cite{zhang2019actionDependentfeedback,weissman2010actionCapacity} is that we assume knowledge of the transmitter's actions at the channel encoder, rather than non-causal state knowledge.

In this work, we establish the secrecy-distortion regions for secure ISAC channels with action-dependent states for physically- and reversely-physically-degraded models.  We first consider the case where channel output feedback is available at the channel encoder alone. Then, we show that the rate regions continue to hold when the channel feedback is available at the action encoder as well. Our achievability proofs leverage the output statistics of random binning (OSRB) framework in \cite{AhlswedeCsiz,OSRBAmin,RenesRenner} to provide strong secrecy. 

In Section~\ref{sec:problem_setting}, we define the secure ISAC with transmitter actions model. In Section~\ref{sec:PSandPOFwithFeedbackToActionEncoder}, we assume that the transmitted message consists of common and secure parts, i.e., only the latter must be kept secret from the eavesdropper \cite{gunlu2023secureISAC,RaviZivPartialSecrecyWTC}. We characterize the secrecy-distortion regions for physically- and reversely-physically-degraded ISAC channels with action-dependent states under this partial secrecy scenario. In Section~\ref{sec:FSandPOFwithFeedbackToActionEncoder}, we simplify the results of Section~\ref{sec:PSandPOFwithFeedbackToActionEncoder} for the case where the entire message should be kept secret from the eavesdropper, i.e., there is no common message, providing full secrecy. Finally, in Section~\ref{sec:binary_example} we illustrate the results by evaluating the secrecy-distortion region under full secrecy of a binary stochastically-degraded example.

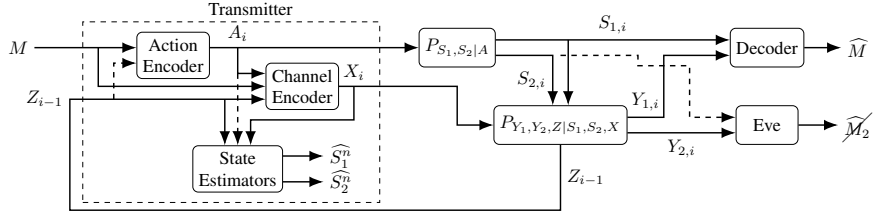
\begin{figure*}[ht]
  \centering
  \resizebox{0.65\linewidth}{!}{
    \begin{tikzpicture}
      \node (a) at (1.5,-.15) [draw,rounded corners = 3pt, minimum width=1.2cm,minimum height=0.75cm, align=center] {Action \\ Encoder};
      \node (b) at (7,0) [draw,rounded corners = 3pt, minimum width=1.2cm,minimum height=0.75cm, align=center] {$P_{S_{1},S_{2}|A}$};
      \node (c) at (4,-.75) [draw, rounded corners= 3pt, minimum width= 1.2cm, minimum height =0.75cm, align=center] {Channel \\ Encoder};
      \node (d) at (9,-1.5) [draw, rounded corners = 3pt, minimum width=1.2cm, minimum height=0.75cm, align=center] {$P_{Y_{1},Y_{2},Z|S_1,S_2,X}$};
      \node (e) at (13,0) [draw, rounded corners = 3pt, minimum width=1.2cm, minimum height=0.75cm, align=center] {Decoder};
      \node (f) at (13,-1.5) [draw, rounded corners = 3pt, minimum width=1.2cm, minimum height=0.75cm, align=center] {Eve};
      \node (g) at (2.75,-2.35) [draw, rounded corners = 3pt, minimum width=1.2cm, minimum height=0.75cm, align=center] {State\\ Estimators};
      \node (h) at (-1.5,0) {$M$};
      \node (i) at (14.75,0) {$\widehat{M}$};
      \node (j) at  (2.625,-1.25) [draw, dashed, minimum width=5.75cm, minimum height=3.5cm, align=center] {};
      \node (k) at (3,.75) {Transmitter};
      \draw[decoration={markings,mark=at position 1 with {\arrow[scale=1.5]{latex}}},
      postaction={decorate}, thick, shorten >=1.4pt] ($(h.east)+(0.0,0)$) -- ($(a.west)-(0,-0.15)$);  
      \node (h1) at (0.05,0) {};
      \node (h2) at (0.05,-.75) {};
      \draw[decoration={markings,mark=at position 1 with {\arrow[scale=1.5]{latex}}},
      postaction={decorate}, thick, shorten >=1.4pt] ($(h1)$) -- ($(h2)$) -- ($(c.west)-(0,0)$);
      \draw[decoration={markings,mark=at position 1 with {\arrow[scale=1.5]{latex}}},
      postaction={decorate}, thick, shorten >=1.4pt] ($(a.east)+(0.0,0.15)$) -- ($(b.west)-(0,0)$);
      \node (a1) at (2.75,0) {};
      \node (a2) at (2.75,-.5) {};
      \draw[decoration={markings,mark=at position 1 with {\arrow[scale=1.5]{latex}}},
      postaction={decorate}, thick, shorten >=1.4pt] ($(a1) + (0,0)$) -- ($(a2)$) node [at start, above] {$A_i$} --  ($(c.west)+ (0,.25)$);
      \draw[decoration={markings,mark=at position 1 with {\arrow[scale=1.5]{latex}}},
      postaction={decorate}, thick, shorten >=1.4pt, dashed] ($(a2) + (0,0)$) -- ($(a2)$) --  ($(g.north)+ (0,0)$);
      \node (c1) at (7,-.75) {};
      \node (c2) at (7,-1.5) {};
      \draw[decoration={markings,mark=at position 1 with {\arrow[scale=1.5]{latex}}},
      postaction={decorate}, thick, shorten >=1.4pt] ($(c.east) + (0,0)$)  -- ($(c1) + (0,0)$) -- ($(c2)$) -- ($(d.west)$);
      \node (c3) at (5,-.75) {};
      \node (c4) at (5,-1.4) {};
      \node (c5) at (3,-1.4) {};
      \draw[decoration={markings,mark=at position 1 with {\arrow[scale=1.5]{latex}}},
      postaction={decorate}, thick, shorten >=1.4pt] ($(c3) + (0,0)$) --  ($(c4)+ (0,0)$) node [at start, above] {$X_i$}  -- ($(c5)$) -- ($(g.north) + (.25,0)$);
      \node (b1) [above of = d, node distance = 1.5cm] {};
      \draw[decoration={markings,mark=at position 1 with {\arrow[scale=1.5]{latex}}},
      postaction={decorate}, thick, shorten >=1.4pt] ($(b.east) - (0,-.15)$) --  ($(e.west) - (0,-.15)$) node [midway, above] {$S_{1,i}$};
      \draw[decoration={markings,mark=at position 1 with {\arrow[scale=1.5]{latex}}},
      postaction={decorate}, thick, shorten >=1.4pt] ($(b1) - (-.15,-.15)$) --  ($(d.north) - (-.15,0)$);
      \draw[decoration={markings,mark=at position 1 with {\arrow[scale=1.5]{latex}}},
      postaction={decorate}, thick, shorten >=1.4pt] ($(b.east) + (0,-.15)$) -- ($(b1) + (-.15,-.15)$) --  ($(d.north) + (-.15,0)$) node [midway, left] {$S_{2,i}$};
      \node (b2) at (10.5,0) {};
      \node (b3) at (10.5,-.35) {};
      \node (f1) at (11.5,-.35) {};
      \node (f2) at (11.5,-1.5) {};
      \draw[decoration={markings,mark=at position 1 with {\arrow[scale=1.5]{latex}}},
      postaction={decorate}, thick, shorten >=1.4pt,dashed] ($(b1) + (0,-.15)$) -- ($(b2) + (0,-.15)$) -- ($(b3)$) -- ($(f1)$) --  ($(f2) + (0,.15)$) --  ($(f.west) + (0,.15)$);
      \node (d1) at (11,-1.5) {};
      \node (d2) at (11,0) {};
      \draw[decoration={markings,mark=at position 1 with {\arrow[scale=1.5]{latex}}},
      postaction={decorate}, thick, shorten >=1.4pt] ($(d.east) - (0,-.15)$) -- ($(d1) - (0,-.15)$) node [midway, above] {$Y_{1,i}$} -- ($(d2) + (0,-.15)$) --  ($(e.west) + (0,-.15)$);
      \draw[decoration={markings,mark=at position 1 with {\arrow[scale=1.5]{latex}}},
      postaction={decorate}, thick, shorten >=1.4pt] ($(d.east) + (0,-.15)$) --  ($(f.west) + (0,-.15)$) node [midway, below] {$Y_{2,i}$};
      \node (d3) at (9,-3.15) {};
      \node (d4) at (-.5,-3.15) {};
      \node (d5) at (-.5,-1) {};
      \node (d6) at (.35,-1) {};
      \node (d7) at (.35,-.3) {};
      \draw[decoration={markings,mark=at position 1 with {\arrow[scale=1.5]{latex}}},
      postaction={decorate}, thick, shorten >=1.4pt] ($(d.south)$) --  ($(d3)$) node [midway, right] {$Z_{i-1}$} -- ($(d4)$) -- ($(d5)$) node [at end, left] {$Z_{i-1}$} -- ($(c.west) + (0,-.25)$);
      \draw[decoration={markings,mark=at position 1 with {\arrow[scale=1.5]{latex}}},
      postaction={decorate}, thick, shorten >=1.4pt,dashed] ($(d6)$) --  ($(d7)$) -- ($(a.west) + (0,-.15)$);
      \node (d8) at (2.5,-1) {};
      \draw[decoration={markings,mark=at position 1 with {\arrow[scale=1.5]{latex}}},
      postaction={decorate}, thick, shorten >=1.4pt] ($(d8)$) -- ($(g.north) + (-.25,0)$);
      \draw[decoration={markings,mark=at position 1 with {\arrow[scale=1.5]{latex}}},
      postaction={decorate}, thick, shorten >=1.4pt] ($(e.east)$) --  ($(i.west)$);
      \node (g3) at (4.75,-2.1) {$\widehat{S_1^n}$};
      \node (g4) at (4.75,-2.6) {$\widehat{S_2^n}$};
      \draw[decoration={markings,mark=at position 1 with {\arrow[scale=1.5]{latex}}},
      postaction={decorate}, thick, shorten >=1.4pt] ($(g.east)+(0,.25)$) --  ($(g3.west)$);
      \draw[decoration={markings,mark=at position 1 with {\arrow[scale=1.5]{latex}}},
      postaction={decorate}, thick, shorten >=1.4pt] ($(g.east)-(0,.25)$) --  ($(g4.west)$);
      \node (f4) at (14.75,-1.5) {$\cancel{\widehat{M}_2}$};
      \draw[decoration={markings,mark=at position 1 with {\arrow[scale=1.5]{latex}}},
      postaction={decorate}, thick, shorten >=1.4pt] ($(f.east)$) -- ($(f4.west)$);
    \end{tikzpicture}
  }
  \vspace*{-0.2cm}
  \caption{Secure ISAC model with action-dependent states under partial secrecy, where $M=(M_1,M_2)$ and only $M_2$ should be kept secret from Eve, for $i~=~[1:n]$. The action, $A_i$, is a random function of $(M,Z_{i-1})$. The channel input, $X_i$, is a random function of $(M,Z_{i-1},A_i)$. We consider ISAC with perfect output feedback, where $Z_{i-1}=(Y_{1,i-1},Y_{2,i-1})$.}\label{fig:SecureStateDependentISACModel}
  \vspace*{-0.3cm}
\end{figure*}

\section{Problem Definition}\label{sec:problem_setting}
We consider the secure ISAC model depicted in Fig.~\ref{fig:SecureStateDependentISACModel}, which consists of a transmitter with  channel encoder, action encoder, and state estimator; a legitimate receiver; and an eavesdropper. The transmitter aims to reliably transmit $M=(M_1,M_2) \in \mathcal{M} = (\mathcal{M}_1\times\mathcal{M}_2)$ to the legitimate receiver over the state-dependent broadcast channel $P_{Y_1Y_2Z|S_1S_2}$ with action-dependent states $(S_1,S_2)$. The transmitter computes the inputs as $X_i=\Enc\big(M,A_i,Z^{i-1}\big)\in\mathcal{X}$ and $A_i=\Enc_{\text{Act}}\big(M,Z^{i-1}\big)\in\mathcal{A}$, where $\Enc(\cdot,\cdot,\cdot)$ and $\Enc_{\text{Act}}(\cdot,\cdot)$ are encoding functions for the channel input and action, respectively, and $Z^{i-1}\in\mathcal{Z}^{i-1}$ is delayed channel output feedback for all $i=[1:n]$. The legitimate receiver observes $Y_{1,i}\in\mathcal{Y}_1$ and $S_{1,i}\in\mathcal{S}_1$ and should be able to form a reliable estimate $\widehat{M} = \Dec(Y_1^n,S_1^n) \in \widehat{\mathcal{M}}$, where $\Dec(\cdot,\cdot)$ is a decoding function. The eavesdropper observes $Y_{2,i}\in\mathcal{Y}_2$ and $S_{2,i}\in\mathcal{S}_2$ and should not be able to recover $M_2$. Finally, the transmitter estimates the states by $\widehat{S_j^n}=\Est_j(A^n,X^n,Z^n)\in\widehat{\mathcal{S}_j^n}$ for $j=1,2$, where $\Est_j(\cdot,\cdot,\cdot)$ is an estimation function. All sets $\mathcal{A},\mathcal{X},\mathcal{Y}_1,\mathcal{Y}_2,\mathcal{S}_1,\mathcal{S}_2,\widehat{\mathcal{S}_1},\widehat{\mathcal{S}_2}$ and $\mathcal{Z}$ are finite.

This channel model abstracts an ISAC scenario, where the transmitter has a multi-functional phased array used for both beamforming the transmitter signals, modeled as an action $A_i$, and observing the resulting reflected waveforms, modeled as the channel output feedback $Z_{i-1}$, from which they derive information about the legitimate receiver's and eavesdropper's channel states $S_{1,i}$ and $S_{2,i}$, respectively. These states can carry information about, e.g., the locations of the receivers.

To simplify the analysis we consider perfect output feedback, i.e. we have $Z_{i-1}=(Y_{1,i-1},Y_{2,i-1})$ for all $i=[2:n]$. While perfect output feedback is an integral part of the achievability proofs, some of the converse results continue to hold for generalized feedback. We next define the strong secrecy-distortion region for the secure ISAC model.

\begin{definition}\label{def:PSPOF}
  \normalfont A secrecy-distortion tuple $(R_{1}, R_{2},D_{1},D_{2})$ is \emph{achievable} under partial secrecy if, for any $\delta\!>\!0$, there exists $n\!\geq\!1$, one channel encoder, one action encoder, one decoder, and two estimators $\Est_j(X^n,A^n,Y_1^n,Y_2^n )=\widehat{S_j^n}$, $j\in\{1,2\}$, such that
  \begin{align}
    & \frac{1}{n}\log |\mathcal{M}_j|\geq R_j -\delta\quad\;\;\;\;\;\text{for } j\!=\!1,2\;\;&&\!\!\!\!\!(\text{rate})\label{eq:rates_cons}\\ 
    &\Pr\big[(M_1,M_2) \neq (\widehat{M}_1,\widehat{M}_2)\big] \leq \delta&&\!\!\!\!\! (\text{reliability})\label{eq:reliability_cons}\\
    &I(M_2;Y^n_2,S_2^n) \leq \delta&&\!\!\!\!\!(\text{strong secrecy})\label{eq:secrecyleakage_cons}\\
    &\mathbb{E}\big[d_j(S_j^n,\widehat{S}_j^n)\big] \!\leq\! D_j\!+\!\delta\;\;\;\;\;\;\text{for } j\!=\!1,2\;\;&&\!\!\!\!\!(\text{distortion})\label{eq:distortion_cons}
  \end{align}
  where $d_j(s^n,\widehat{s}^n)=\frac{1}{n}\sum_{i=1}^nd_j(s_i,\widehat{s}_i)$ for $j\!=\!1,2$ are bounded per-letter distortion metrics. 
  
  The secrecy-distortion region $\mathcal{R}_{\textnormal{PS,Act}}$ is the closure of the set of all achievable tuples under partial secrecy and perfect output feedback. \hfill $\lozenge$
\end{definition}

\begin{remark}\label{rem:secrecyconstraintwithoutcond}
  \normalfont In \cite{gunlu2023secureISAC}, the independence of the message and the states allowed the strong secrecy condition (\ref{eq:secrecyleakage_cons}) to be simplified to $I(M_2;Y^n_2|S_2^n) \leq \delta$. In our model, the action introduces dependence between the message and states, making $I(M_2;S_2^n)\!\neq\!0$, invalidating the simplification $I(M_2;Y^n_2,S_2^n)\!=\!I(M_2;Y^n_2|S_2^n)$. Moreover, unlike in \cite{gunlu2023secureISAC}, the channel input $X_i$ is not independent of the channel states $(S_{1,i},S_{2,i})$ in our model. 
\end{remark}

We now define the physically degraded model, see \cite{MariMichelleGJournal}.
\begin{definition}\label{def:physicallydegraded}
  \normalfont An ISAC channel as depicted in Fig.~\ref{fig:SecureStateDependentISACModel} is \emph{physically-degraded} if we have
  \begin{align}
    P_{AXY_1Y_2S_1S_2}&=P_{AX}P_{Y_1S_1Y_2S_2|AX}\nonumber\\
    &=P_{AX}P_{S_1|A}P_{Y_1|S_1X}P_{Y_2S_2|Y_1S_1}\label{eq:physicaldegradedcond}.
  \end{align}
  Similarly, it is \emph{reversely-physically-degraded} if the degradation order is swapped such that
  \begin{align}
    P_{AXY_1Y_2S_1S_2}&=P_{AX}P_{Y_1S_1Y_2S_2|AX}\nonumber\\
    &=P_{AX}P_{S_2|A}P_{Y_2|S_2X}P_{Y_1S_1|Y_2S_2}.\label{eq:reversephysicallydegradedcond}
  \end{align}\hfill $\lozenge$
\end{definition}

\section{ISAC with Action-dependent States Under Partial Secrecy}\label{sec:PSandPOFwithFeedbackToActionEncoder}
We provide the strong secrecy-distortion regions for the physically- and reversely-physically-degraded secure ISAC channels with transmitter actions. Proof of Theorem~\ref{theo:PSPOFPhysDegraded} is given in Appendix~\ref{app:thm1}, a proof sketch is provided here.


\begin{theorem}\label{theo:PSPOFPhysDegraded}
  {\normalfont(Physically-degraded):} For a physically-degraded ISAC channel with strictly causal feedback available at the action and channel encoders, $\mathcal{R}_{\textnormal{PS,Act}}$ is the union over all joint distributions $P_{VAX}$ of the rate tuples $(R_{1}, R_{2},D_1,D_2)$ satisfying
  \begin{align}
    &R_{1}\leq I(V;Y_1,S_1)\label{eq:achPD1}\\
    &R_{2}\leq \min\{R_{2}^{\prime}, (I(V;Y_1,S_1)-R_1)\}\label{eq:achPD2}\\
    & D_j\geq \mathbb{E}[d_j(S_j,\widehat{S}_j))]\qquad\qquad  \text{for }j=1,2\label{eq:achdistortion1and2}
  \end{align}
  where we have
  \begin{align}
    &P_{VAXY_1Y_2S_1S_2}\!=\!P_{V|AX}P_{AX}P_{S_1|A}P_{Y_1|S_1X}P_{Y_2S_2|S_1Y_1}\label{eq:jointprobPD},\\
    &R_{2}^{\prime}=H(Y_1,S_1|Y_2,S_2)-H(S_1|Y_1,Y_2,S_2,V)\label{eq:R2primedef}
  \end{align}
  and one can use the deterministic per-letter estimators $\Est_j(a,x,y_1,y_2)=~\hat{s}_j$ for $j=1,2$, such that 
  \begin{align}
    &\Est_j(a,x,y_1,y_2)\nonumber\\
    &=\mathop{\textnormal{argmin}}_{\tilde{s}\in\widehat{\mathcal{S}}_j} \sum_{s_j\in\mathcal{S}_j}P_{S_j|AXY_1Y_2}(s_j|a,x,y_1,y_2)\; d_j(s_j,\tilde{s}).\label{eq:deterministicest} 
  \end{align}
  One can also bound $|\mathcal{V}|$ as
  \begin{align}
    &\min\{|\mathcal{X}|\!\cdot\!|\mathcal{A}|,\;|\mathcal{Y}_1|\!\cdot\!|\mathcal{S}_1|,\;|\mathcal{Y}_2|\!\cdot\!|\mathcal{S}_2|\}\!+\!1.\label{eq:cardVforPS}
  \end{align}
\end{theorem}

\begin{IEEEproof}[Proof Sketch] 
  For the achievability proof, we leverage a block-Markov coding scheme with $B\!\geq\!2$ blocks, each with $n$ channel uses, wherein block $b\in[2:B]$, where $[1:n]=\{1,2,\dots,n\}$ a part of the common message is hidden with a key derived from block $b-1$. We next show reliability and security guarantees for a single block.

  Fix $P_{VAX}$ such that there exist per-letter estimators $\Est_j(a,x,y_1,y_2)=~\hat{s}_j$ that satisfy $\mathbb{E}[d_j(S_j,\widehat{S}_j))]\!\leq\!D_j + \epsilon_n$
  for $j=1,2$, where $\epsilon_n\!\geq0$ and $\epsilon_n\rightarrow0$ when $n\rightarrow\infty$. Let $(V^n,A^n,X^n,Y_1^n,S_1^n,Y_2^n,S_2^n)$ be generated independently and identically distributed (i.i.d.) according to \eqref{eq:jointprobPD}. To the sequence $v^n$, we uniformly and independently assign random bin indices $W_{\text{v}_1}\in[1:2^{nR_{\text{v}_1}}]$, $W_{\text{v}_2}\in[1:2^{nR_{\text{v}_2}}]$, $L_{\text{v}}\in[1:2^{n\overline{R}_{\text{v}}}]$, and $F_{\text{v}}\in[1:2^{n\widetilde{R}_{\text{v}}}]$. We also assign to $y_1^n$ a random bin index $L_{\text{y}_1}\in[1:2^{n\overline{R}_{\text{y}_1}}]$, and we set $\overline{R}_{\text{y}_1}=\overline{R}_{\text{v}}$. We next choose $F=F_{\text{v}}$ and $M=(M_1,M_2)$ with $M_1=W_{\text{v}_1}$ and $M_2=(W_{\text{v}_1},L_{\text{y}_1}\oplus L_{\text{v}})$, where $\oplus$ is the one-time padding operation. Conceptually, $M$ represents the message $(M_1,M_2)$, whereas $F$ represents the randomness that defines the codebook known to all parties.

  The rate conditions on the random bin indices which guarantee reliability and secrecy are as follows.
  
  \emph{Reliability}: Using a Slepian-Wolf \cite{SW} decoder, $V^n$ can be reliably recovered from $(Y_1^n,S_1^n,F)$ if~\cite[Lemma 1]{OSRBAmin}
  \begin{align}
      \widetilde{R}_{\text{v}}\!>\!H(V|Y_1,S_1)\label{eq:reliabilityPD}.
  \end{align}
  \emph{Secrecy}: Using privacy amplification~\cite[Theorem 1]{OSRBAmin}, $W_{\text{v}_2}$ and $F$ become almost independent of $(Y_2^n,S_2^n)$ and uniformly distributed if
  \begin{align}
      R_{\text{v}_2} + \widetilde{R}_{\text{v}}\!<\!H(V|Y_2,S_2). \label{eq:secPDwtr} 
  \end{align}
  Similarly, $L_{\text{y}_1}$ becomes almost independent of $(Y_2^n,S_2^n,V^n)$ and uniformly distributed if
  \begin{align}
      \overline{R}_{\text{y}_1} = \overline{R}_{\text{v}}\!<\!H(Y_1|Y_2,S_2,V). \label{eq:secPDkey}
  \end{align}
  Then, it follows that $L_{\text{y}_1}\oplus L_{\text{v}}$ is also almost independent of $(Y_2^n,S_2^n,V^n)$. Thus, $M_2$ is almost independent of the eavesdropper's observations, yielding the strong secrecy condition. 
  
  Note that all $(v^n,x^n,y_1^n,y_2^n,s_1^n,s_2^n)$ tuples are in the jointly typical set with high probability. Using the law of total expectation to bounded distortion metrics and the typical average lemma \cite[pp.~26]{Elgamalbook}, proves that the distortion constraints (\ref{eq:distortion_cons}) are satisfied. The sufficiency of the deterministic estimators follows similarly to \cite[Lemma 1]{MariMichelleGJournal}.

  By \cite[Theorem 1]{OSRBAmin}, the distribution induced by the source coding scheme is arbitrarily close in variational distance to the distribution induced by channel coding scheme when
  \begin{align}
      \widetilde{R}_{\text{v}} + R_{\text{v}_1} + R_{\text{v}_2} + \overline{R}_{\text{v}}\!<H(V).\label{eq:ABdualityPD}
  \end{align}
  Applying Fourier-Motzkin elimination \cite{FMEbook} to \eqref{eq:reliabilityPD}-\eqref{eq:ABdualityPD} and setting $R_1\!=\!R_{\text{v}_1}$ and $R_2\!=\!R_{\text{v}_1}+\overbar{R}_{\text{v}}$, we recover \eqref{eq:achPD1} and
  \begin{align}
      R_2\!\leq\!\min\big\{R_{\text{sec}},I(V;Y_1,S_1)-R_1\big\} \label{eq:R2general}
  \end{align}
  where we have $R_{\text{sec}}$ is equal to
  \begin{align}
       \big[H(V|Y_2,S_2) - H(V|Y_1,S_1)\big]^+ +H(Y_1|Y_2,S_2,V)  \label{eq:R2sec}
  \end{align}
  with $[b]^+=\max\{b,0\}$ for $b\in\mathbb{R}$. 
  
  The channel with no degradedness assumptions conforms to the Markov chain
  \begin{align}
      V-(X,A)-(Y_1,S_1,Y_2,S_2) \label{eq:channelMC}.
  \end{align}
  Combining \eqref{eq:physicaldegradedcond} and \eqref{eq:channelMC} provides the Markov chain
  \begin{align}
      V-(X,A)-(Y_1,S_1)-(Y_2,S_2). \label{eq:pdmc}
  \end{align} 
  Applying \eqref{eq:pdmc} to \eqref{eq:R2sec} results in \eqref{eq:achPD2}.

  As the secret key cannot be used in the same block in which it is generated, we turn to the block-Markov coding scheme as described above, and use the secret $L_{\text{y}_1}$ in block $b-1$ as a key to secure $L_{\text{v}}$ in block $b$. In the block-Markov coding scheme, there is no secret message $M_2$ sent in the first block. A union bound on the probability of making a decoding error in each block shows that the reliability condition \eqref{eq:reliability_cons} is asymptotically satisfied. The proof that the secrecy constraint is unaffected across all blocks follows similarly to the proof of \cite[Proposition 1]{gunlu2023secureISAC}.

  For the converse proof, assume that for some $\delta_n>0$, with $\delta_n\rightarrow 0$ as $n\rightarrow\infty$, there exist an action encoder, channel encoder, decoder, and estimators such that \eqref{eq:rates_cons}-\eqref{eq:distortion_cons} are satisfied for $(R_1,R_2,D_1,D_2)$. We define $V_{i}\triangleq (M_1,M_2,Y^{i-1}_{1},S_1^{i-1},Y^{i-1}_{2},S_2^{i-1})$ such that $V_i-(A_i,X_i)-(Y_{1,i},S_{1,i},Y_{2,i},S_{2,i})$ forms a Markov chain for all $i\in[1:~n]$.

  \emph{Bound on $R_1$}: We have 
  \begin{align}
      & nR_1 \overset{(a)}{\leq} \sum\limits_{i=1}^n\big[H\big(Y_{1,i},S_{1,i}|Y_1^{i-1},S_1^{i-1}\big)+\epsilon_n\nonumber\\
           & \qquad\qquad\qquad -H\big(Y_{1,i},S_{1,i}|M,Y_1^{i-1},S_1^{i-1},Y_2^{i-1},S_2^{i-1}\big) \big]\nonumber\\
           & \overset{(b)}{=} \sum\limits_{i=1}^n\big[I\big(V_i;Y_{1,i},S_{1,i}\big) + \epsilon_n\big] \label{eq:R1BoundConv}
  \end{align}
  where $(a)$ follows from Fano's inequality \cite{CoverandThomas} with $\epsilon_n\rightarrow0$ when $\delta_n\rightarrow0$ and $(b)$ follows from the definition of $V_i$.

  \emph{Bound on $R_1+R_2$}: We have
  \begin{align}
      n(R_1+R_2) & \leq \sum\limits_{i=1}^n\big[I\big(V_i;Y_{1,i},S_{1,i}\big) + \epsilon_n\big]
  \end{align}
  which follows similarly to \eqref{eq:R1BoundConv}.

  \emph{Bound on $R_2$}: We obtain
  \begin{align}
      & nR_2 \overset{(a)}{\leq} I(M_2;Y_1^n,S_1^n,Y_2^n,S_2^n) + n\epsilon_n \nonumber\\
      & = H(Y_1^n,S_1^n|Y_2^n,S_2^n) + I(M_2;Y_2^n,S_2^n) \nonumber\\
      &\quad\quad- H(Y_1^n,S_1^n|M_2,Y_2^n,S_2^n) + n\epsilon_n \nonumber\\
      & \overset{(b)}{\leq} H(Y_1^n,S_1^n|Y_2^n,S_2^n) + \delta_n \nonumber\\
      &\quad\quad- H(S_1^n|M,Y_1^n,Y_2^n,S_2^n) +n\epsilon_n \nonumber\nonumber\\
      & \leq \sum\limits_{i=1}^n\big[H(Y_{1,i},S_{1,i}|Y_{2,i},S_{2,i})\nonumber\\
      &\qquad\qquad- H(S_{1,i}|M,Y_1^n,S_1^{i-1},Y_2^n,S_2^n)\big] + n\epsilon_n + \delta_n\nonumber
            \end{align}
      \begin{align}
      & \overset{(c)}{=} \sum\limits_{i=1}^n\big[H(Y_{1,i},S_{1,i}|Y_{2,i},S_{2,i})\nonumber\\
      & \qquad\qquad- H(S_{1,i}|Y_{1,i},Y_{2,i},S_{2,i},V_i) + \epsilon_n + \frac{1}{n}\delta_n\big] \label{eq:R2BoundConv}
  \end{align}
  where $(a)$ follows from Fano's inequality similar to the bound on $R_1$, $(b)$ follows by \eqref{eq:secrecyleakage_cons}, and $(c)$ is a consequence of the definition of $V_i$ and the Markov chain 
  \begin{align*}
    (Y_{1,i+1}^n,Y_{2,i+1}^n,S_{2,i+1}^n)-(M,Y_1^i,S_1^{i-1},Y_2^i,S_2^i)-S_{1,i}.
  \end{align*}

  The deterministic estimators, for j=1,2, follow from
  \begin{align}
      D_j+\delta_n \geq \mathbb{E}\big[d_j(S_j^n,\widehat{S}_j^n)\big]=\frac{1}{n}\sum\limits_{i=1}^n\mathbb{E}\big[d_j(S_{j,i},\widehat{S}_{j,i})\big].
  \end{align}

  Next, we introduce a time-sharing random variable $Q$ distributed uniformly on $[1:n]$ which is independent of all other random variables and defining $X=X_Q$, $A=A_Q$, $Y_1=Y_{1,Q}$, $S_1=S_{1,Q}$, $Y_1=Y_{1,Q}$, $S_2=S_{2,Q}$, and $V=(V_Q,Q)$ so that $V-(X,A)-(Y_1,S_1,Y_2,S_2)$ forms a Markov chain. Letting $\delta_n\rightarrow0$ gives \eqref{eq:achPD1}-\eqref{eq:achdistortion1and2}.
  
  The proof of the cardinality bound for $V$ follows from the support lemma \cite[Lemma 15.4]{CsiszarKornerbook2011}.
\end{IEEEproof}

\begin{remark} 
    \normalfont Note that the converse proof of Theorem~\ref{theo:PSPOFPhysDegraded} does not use the physical degradedness of the channel. \label{remark:general_converse}
\end{remark}

We next provide the secrecy-distortion region for reversely-physically-degraded ISAC channels. Proof of Theorem~\ref{theo:PSPOFRevPhysDegraded} is given in Appendix~\ref{app:thm2}, a proof sketch is provided here.

\begin{theorem}\label{theo:PSPOFRevPhysDegraded}
  {\normalfont(Reversely-physically-degraded):} For a reversely-physically-degraded ISAC channel with strictly causal feedback available at the action and channel encoders, $\mathcal{R}_{\textnormal{PS,Act}}$ is the union over all joint distributions $P_{VAX}$ of the rate tuples $(R_{1}, R_{2},D_1,D_2)$ satisfying the rate constraint \eqref{eq:achPD1}, the distortion constraints in \eqref{eq:achdistortion1and2}, and
  \begin{align}
    &R_{2}\leq \min\{H(Y_1|Y_2,S_2), (I(V;Y_1,S_1)-R_1)\}\label{eq:achPDR2}
  \end{align}
  where
  \begin{align}
    &P_{VAXY_1Y_2S_1S_2} \!=\!P_{V|AX}P_{AX}P_{S_2|A}P_{Y_2|S_2X}P_{Y_1S_1|S_2Y_2}\label{eq:jointprobRevPD}
  \end{align}
  using the estimators in \eqref{eq:deterministicest} with $|\mathcal{V}|$ bounded above by \eqref{eq:cardVforPS}.
\end{theorem}


\begin{IEEEproof}[Proof Sketch]
  The proof of Theorem~\ref{theo:PSPOFRevPhysDegraded} follows similarly to that of Theorem~\ref{theo:PSPOFPhysDegraded}, so we highlight the differences below.

  Combining the definition of reversely-physical degradation in \eqref{eq:reversephysicallydegradedcond} with the channel Markov chain \eqref{eq:channelMC} gives the Markov chain
  \begin{align}
      V-(X,A)-(Y_1,S_1)-(Y_2,S_2). \label{eq:rpdmc}
  \end{align}
  
  For the achievability proof, random binning applied to $(V^n,A^n,X^n,Y_1^n,S_1^n,Y_2^n,S_2^n)$ generated i.i.d. according to \eqref{eq:jointprobRevPD} in place of \eqref{eq:jointprobPD} yields the conditions \eqref{eq:achPD1}, \eqref{eq:achdistortion1and2}, and \eqref{eq:cardVforPS}. The intermediate steps \eqref{eq:R2general} and \eqref{eq:R2sec} also follow similarly, but the simplification differs because we use \eqref{eq:rpdmc} in place of \eqref{eq:pdmc}, resulting in \eqref{eq:achPDR2}.

  For the converse proof, the bounds on $R_1$ and $(R_1+R_2)$ and the distortion constraints follow as in the proof of Theorem~\ref{theo:PSPOFPhysDegraded}. Noting Remark~\ref{remark:general_converse}, we can directly apply the reverse degraded condition in \eqref{eq:rpdmc} to \eqref{eq:achPD2}, which simplifies to \eqref{eq:achPDR2}.
\end{IEEEproof}


For physically-degraded or reversely-physically-degraded secure ISAC channels with strictly causal feedback available at both the action and channel encoders, $\mathcal{R}_{\textnormal{PS,Act}}$ stays the same as given in Theorem~\ref{theo:PSPOFPhysDegraded} and \ref{theo:PSPOFRevPhysDegraded}. This follows because the Markov chains used in the proofs of Theorems~\ref{theo:PSPOFPhysDegraded} and \ref{theo:PSPOFRevPhysDegraded} continue to hold when we introduce dependence on the channel feedback for the actions. This parallels the results on channels with causal knowledge of action-dependent states in \cite{weissman2010actionCapacity}. The corresponding model in \cite{weissman2010actionCapacity} assumes that $S_i$ is available at the transmitter, while we only assume that $A_i$, generated by the action encoder before channel use $i$, is available for use in computing the channel input.

\section{ISAC with Action-dependent States Under Full Secrecy}\label{sec:FSandPOFwithFeedbackToActionEncoder}
We next consider secure action-dependent ISAC with full secrecy, i.e., $M=M_2$, or $M_1=\emptyset$, which is the case when the entire message should be kept secret from the eavesdropper.

\begin{definition}\label{def:fullsecrecyPOF}
  \normalfont A secrecy-distortion tuple $(R,D_{1},D_{2})$ is \emph{achievable} under full secrecy if, for any $\delta\!>\!0$, there exist $n\!\geq\!1$, one channel encoder, one action encoder, one decoder, and two estimators $\Est_j(X^n,A^n,Y_1^n,Y_2^n )=\widehat{S_j^n}$, $j\in\{1,2\}$, such that
  \begin{align}
    & \frac{1}{n}\log |\mathcal{M}|\geq R -\delta&&\!\!\!\!\!(\text{rates})\label{eq:FSrates_cons}\\ 
    &\Pr\big[M \neq \widehat{M}\big] \leq \delta&&\!\!\!\!\! (\text{reliability})\label{eq:FSreliability_cons}\\
    &I(M;Y^n_2,S_2^n) \leq \delta&&\!\!\!\!\!(\text{strong secrecy})\label{eq:FSsecrecyleakage_cons}
  \end{align}
  and the distortion constraints \eqref{eq:distortion_cons} are satisfied. 
  
  The secrecy-distortion region $\mathcal{R}_{\textnormal{Act}}$ is the closure of the set of all achievable tuples under full secrecy and perfect output feedback. \hfill $\lozenge$
\end{definition}


We next give the rate region under full secrecy for the physically-degraded channels. Proof of Theorem~\ref{theo:PDfullsecrecy} is given in Appendix~\ref{app:thm3}, a proof sketch is provided here.

\begin{theorem} (Physically-degraded) \label{theo:PDfullsecrecy}
    For a physically-degraded ISAC channel with strictly causal feedback available at the action and channel encoders, $\mathcal{R}_{Act}$ is the union over all joint distributions $P_{AX}$ of the rate tuples $(R,D_1,D_2)$ satisfying 
      \begin{align}
    &R\leq \min\{H(Y_1,S_1|Y_2,S_2)-H(S_1|Y_1,Y_2,S_2,X,A),\nonumber\\
    &\quad\quad\quad\quad\quad I(X,A;Y_1,S_1)\}\label{eq:FSachPD2}
  \end{align}
  and the distortion constraints in \eqref{eq:achdistortion1and2} using the estimators in \eqref{eq:deterministicest}, with distribution \eqref{eq:physicaldegradedcond}.
\end{theorem}

\begin{IEEEproof}[Proof Sketch]
    The achievability follows by removing $V^n$, fixing $P_{AX}$, generating a tuple of random variables i.i.d. according to \eqref{eq:physicaldegradedcond}, and performing the random binning on $(a^n,x^n)$, removing $W_{V_1}$, replacing $W_{V_2}$ with $W_{AX}\in\textnormal{Unif}[1:2^{nR_{ax}}]$, and replacing $L_{V}$ with $L_{AX}\in\textnormal{Unif}[1:2^{n\overbar{R}_{ax}}]$, and performing the rate simplifications as in the proof of Theorem~\ref{theo:PSPOFPhysDegraded}.

    We next outline the two bounds on $R$.
    \begin{align}
        nR&\overset{(a)}{\leq}\sum\limits_{i=1}^n\big[H(Y_{1,i},S_{1,i})\nonumber\\
        &\qquad\qquad-H(Y_{1,i},S_{1,i}|Y_1^{i-1},S_1^{i-1},M,X_i,A_i)\big] + n\epsilon_n\nonumber
        \end{align}
        \begin{align}
        &\overset{(b)}{=}\sum\limits_{i=1}^nI(X_i,A_i;Y_{1,i},S_{1,i})+n\epsilon_n
    \end{align}
    where $(a)$ follows from Fano's inequality, where $\epsilon_n\rightarrow0$ when $\delta_n\rightarrow0$, and $(b)$ is a result of the Markov chain 
    \begin{align*}
        (Y_{1,i},S_{1,i})-(X_i,A_i)-(Y_1^{i-1},S_1^{i-1},M).
    \end{align*}

    Similar to the bound on $R_2$ in \eqref{eq:R2BoundConv}, we have
    \begin{align}
        & nR \overset{(a)}{\leq}\sum\limits_{i=1}^n\big[H(Y_{1,i},S_{1,i}|Y_{2,i},S_{2,i})+\epsilon_n\nonumber\\
        &\qquad\qquad\quad -H(S_{1,i}|Y_{1,i}^n,Y_{2,i}^n,S_{2,i}^n,M,S_1^{i-1},X_i,A_i)\big] + \delta_n\nonumber\\
        & \overset{(b)}{=}\sum\limits_{i=1}^n\big[H(Y_{1,i},S_{1,i}|Y_{2,i},S_{2,i})+\epsilon_n\nonumber\\
        &\qquad\qquad\quad -H(S_{1,i}|Y_{1,i},Y_{2,i},S_{2,i},X_i,A_i)\big] +\delta_n
    \end{align}
    where $(a)$ follows by Fano's inequality with $\epsilon_n\rightarrow0$ when $\delta_n\rightarrow0$, and the secrecy constraint \eqref{eq:FSsecrecyleakage_cons} and $(b)$ follows by application of the Markov chain 
    \begin{align*}
        S_{1,i}\!-\!(Y_{1,i},Y_{2,i},S_{2,i},X_i,A_i)\!-\!(Y_1^{n\setminus i},Y_2^{n\setminus i},S_2^{n\setminus i},M,S_1^{i-1}).
    \end{align*}
    Introducing a time-sharing random variable, applying the distortion bounds  in \eqref{eq:distortion_cons}, and letting $\delta_n\rightarrow0$ gives the result.
\end{IEEEproof}

We next provide the rate region under full secrecy for the reversely-physically-degraded channels. Proof of Theorem~\ref{theo:RPDfullsecrecy} is given in Appendix~\ref{app:thm4}, a proof sketch is provided here.

\begin{theorem} (Reversely-physically-degraded) \label{theo:RPDfullsecrecy}
    For a reversely-physically-degraded ISAC channel with delayed channel feedback available at the action and channel encoders, $\mathcal{R}_{Act}$ is the union over all joint distributions $P_{AX}$ of the rate tuples $(R,D_1,D_2)$ satisfying 
      \begin{align}
    &R\leq \min\{H(Y_1|Y_2,S_2), I(X,A;Y_1,S_1)\}\label{eq:FSachRPD2}
  \end{align}
  and the distortion constraints in \eqref{eq:achdistortion1and2} using the estimators in \eqref{eq:deterministicest}, with distribution \eqref{eq:reversephysicallydegradedcond}.
\end{theorem}

\begin{IEEEproof}[Proof Sketch]
    The achievability proof follows from the steps in the proof of Theorem~\ref{theo:PDfullsecrecy} and performing the rate simplifications as in the proof of Theorem~\ref{theo:PSPOFRevPhysDegraded}. The converse proof follows by simplifying the results of the converse results of Theorem~\ref{theo:PDfullsecrecy} with the reverse-physical-degradation, as in the proof of the converse for Theorem~\ref{theo:PSPOFRevPhysDegraded}.
\end{IEEEproof}

\section{Binary Noiseless Secure ISAC Channels with Action-dependent States} \label{sec:binary_example}
Suppose a secure ISAC scenario with perfect output feedback, full secrecy, and action-dependent multiplicative Bernoulli states. The input, output, and action alphabets are binary. Specifically, we have
\begin{align}
    Y_1 = S_1\cdot X,\qquad\qquad Y_2=S_2\cdot X \label{eq:ExampleChannel},
\end{align}
and
\begin{alignat}{2}
    &P_{S_1S_2|A}(0,0|0) = \lambda,                 \;\;\;\;\; &&P_{S_1S_2|A}(1,0|0) = (1\!-\!\lambda)(1-\alpha),\nonumber
        \end{alignat}
    \begin{alignat}{2}
    &P_{S_1S_2|A}(0,1|0) = 0,                        &&P_{S_1S_2|A}(1,1|0) = (1\!-\!\lambda),\alpha\nonumber\\
    &P_{S_1S_2|A}(0,0|1) = 1-\lambda,\;\;\;\;                &&P_{S_1S_2|A}(1,0|1) = \lambda(1-\alpha),\nonumber\\
    &P_{S_1S_2|A}(0,1|1) = 0,                        &&P_{S_1S_2|A}(1,1|1) = \lambda\alpha,\label{eq:ExampleConditionalStateDistribution}
\end{alignat}
and 
\begin{alignat}{2}
    &P_{XA}(0,0)  = (1-p)(1-q),  \quad &&P_{XA}(1,0)=pq,\nonumber\\
    &P_{XA}(0,1)  = (1-p)q,      \quad&& P_{XA}(1,1) = p(1-q)
\end{alignat} 
with $P_{XAS_1S_2}=P_{XA}P_{S_1S_2|A}$ for fixed $\lambda,\alpha\in[0,1]$. This ISAC channel is stochastically-degraded, i.e., there exists a marginal probability distribution so that the ISAC channel can be represented as \eqref{eq:physicaldegradedcond}. The constraints \eqref{eq:FSrates_cons}-\eqref{eq:FSsecrecyleakage_cons}, and \eqref{eq:distortion_cons} in Definition~\ref{def:fullsecrecyPOF} only depend on the marginal distributions of $(X,A,Y_1,S_1)$ and $(X,A,Y_2,S_2)$ when per-letter estimators of the form $\mathsf{Est}_j(x,a,y_j)$ are used for $j=1,2$, so the secrecy-distortion region in Theorem~\ref{theo:PDfullsecrecy} is also valid for stochastically-degraded secure ISAC channels. 

Define $p\ast q=(1-p)q+p(1-q)$ and $H_b(x)=-x\log x- (1-x)\log (1-x)$. $X\sim\text{Bern}(p)$ represents a Bernoulli random variable $X$ with probability $p$ of success. Proof of Lemma~\ref{lem:BinaryExample} is given in Appendix~\ref{app:lem1}.

\begin{lemma} \label{lem:BinaryExample}
    The strong secrecy-distortion region $R_{Act}$ for a binary ISAC channel with transmitter actions and multiplicative Bernoulli states, characterized by \eqref{eq:ExampleConditionalStateDistribution} for fixed $\lambda,\alpha\in[0,1]$ with Hamming distortion metrics is the union over all $p,q\in[0,1]$, where $X\sim\textnormal{Bern}(p)$ and $A\sim\textnormal{Bern}(p\ast q)$, of the rate tuples $(R,D_1,D_2)$ satisfying
    \begin{align}
        R&\leq\min\Bigg\{\Bigg(\big(1-(1-p)q(1-\alpha\lambda\big)H_b\bigg(\frac{1-\lambda}{1-\alpha\lambda}\bigg)\nonumber\\ 
        &\qquad\qquad +(1-\alpha)(1-\lambda\ast p\ast q)H_b\bigg(\frac{p(q\ast\lambda)}{1-\lambda\ast p\ast q}\bigg)\nonumber\\ 
        &\qquad\qquad - (1\!-\!p)(1\!-\!q)(1\!-\!\alpha\!+\!\alpha\lambda) H_b\bigg(\frac{\lambda}{1-\alpha + \lambda\alpha}\bigg)\Bigg),\nonumber\\
        &\Bigg((1-\lambda)(p\ast q)\log(1-\lambda)+\lambda(1-p\ast q)\log\lambda \nonumber\\
        &-(1-\lambda)\big((1-p)(1-q)\log((1-p)(1-q)) -pq\log pq\big)\nonumber\\
        &-\lambda\big((1-p)q\log((1-p)q)+p(1-q)log(p(1-q))\big)\nonumber\\
        &-(\lambda\ast p\ast q)\log(\lambda\ast p\ast q)\!+\! H_b\bigg(\!\frac{q\lambda}{1\!-\!q\!\ast\!\lambda}\!\bigg)\!(1\!-\!p)(1\!-\!q\!\ast\!\lambda) \nonumber\\
        & \!+\! H_b\bigg(\!\frac{q(1\!-\!\lambda)}{q\ast\lambda}\!\bigg)\!p(q\ast\lambda)\Bigg)\Bigg\} \label{eq:lemRcondition}\\
        D_1&\geq (1-p)\min\{\lambda,1-\lambda\},\label{eq:lemD1}\\
        D_2&\geq (1-p)\big((1-q)\min\{1-\alpha+\alpha\lambda,\alpha-\alpha\lambda\}\nonumber\\
        &\qquad\quad\qquad+q\min\{1-\alpha\lambda,\alpha\lambda\}\big)\label{eq:lemD2}.
    \end{align}
\end{lemma}


\section*{Acknowledgment}
This work was supported in part by the by the U.S. Department of Transportation under Grant 69A3552348327 for the CARMEN+ University Transportation Center, ZENITH Research and Leadership Career Development Fund, Chalmers Transport Area of Advance, and the ELLIIT funding.

\clearpage
\newpage
\bibliographystyle{IEEEtran}
\bibliography{references}

\begin{thebibliography}{10}
\providecommand{\url}[1]{#1}
\csname url@samestyle\endcsname
\providecommand{\newblock}{\relax}
\providecommand{\bibinfo}[2]{#2}
\providecommand{\BIBentrySTDinterwordspacing}{\spaceskip=0pt\relax}
\providecommand{\BIBentryALTinterwordstretchfactor}{4}
\providecommand{\BIBentryALTinterwordspacing}{\spaceskip=\fontdimen2\font plus
\BIBentryALTinterwordstretchfactor\fontdimen3\font minus \fontdimen4\font\relax}
\providecommand{\BIBforeignlanguage}[2]{{%
\expandafter\ifx\csname l@#1\endcsname\relax
\typeout{** WARNING: IEEEtran.bst: No hyphenation pattern has been}%
\typeout{** loaded for the language `#1'. Using the pattern for}%
\typeout{** the default language instead.}%
\else
\language=\csname l@#1\endcsname
\fi
#2}}
\providecommand{\BIBdecl}{\relax}
\BIBdecl

\bibitem{NokiaGuysJCASTutorial}
T.~Wild, V.~Braun, and H.~Viswanathan, ``Joint design of communication and sensing for beyond {5G} and {6G} systems,'' \emph{{IEEE} {A}ccess}, vol.~9, pp. 30\,845--30\,857, Feb. 2021.

\bibitem{MariMichelleGJournal}
M.~Ahmadipour, M.~Kobayashi, M.~Wigger, and G.~Caire, ``An information-theoretic approach to joint sensing and communication,'' \emph{{IEEE} Trans. Inf. Theory}, May 2022.

\bibitem{liu2022isac}
F.~Liu, Y.~Cui, C.~Masouros, J.~Xu, T.~X. Han, Y.~C. Eldar, and S.~Buzzi, ``Integrated sensing and communications: Toward dual-functional wireless networks for 6g and beyond,'' \emph{IEEE journal on selected areas in communications}, vol.~40, no.~6, pp. 1728--1767, 2022.

\bibitem{tekin2005secureGMAWTC}
E.~Tekin, S.~Serbetli, and A.~Yener, ``On secure signaling for the {Gaussian} multiple access wire-tap channel,'' in \emph{Proc. 2005 Asilomar Conf. On Signals, Systems, and Computers}, 2005, pp. 1747--1751.

\bibitem{ourbinaryAWGNISAC}
O.~Günlü, M.~Bloch, R.~F. Schaefer, and A.~Yener, ``Secure integrated sensing and communication for binary input additive white {Gaussian} noise channels,'' in \emph{{IEEE} Int. Symp. Joint Commun. Sensing}, Seefeld, Austria, Mar. 2023, pp. 1--6.

\bibitem{JCASwithSecurityTutorial}
Z.~Wei, F.~Liu, C.~Masouros, N.~Su, and A.~P. Petropulu, ``Toward multi-functional 6g wireless networks: Integrating sensing, communication, and security,'' \emph{IEEE Communications Magazine}, vol.~60, no.~4, pp. 65--71, 2022.

\bibitem{gunlu2023secureISAC}
O.~G{\"u}nl{\"u}, M.~R. Bloch, R.~F. Schaefer, and A.~Yener, ``Secure integrated sensing and communication,'' \emph{IEEE Journal on Selected Areas in Information Theory}, 2023.

\bibitem{ren2023beamSecureISAC}
Z.~Ren, L.~Qiu, J.~Xu, and D.~W.~K. Ng, ``Robust transmit beamforming for secure integrated sensing and communication,'' \emph{IEEE Transactions on Communications}, 2023.

\bibitem{bazzi2024secureFullDuplex}
A.~Bazzi and M.~Chafii, ``Secure full duplex integrated sensing and communications,'' \emph{IEEE Transactions on Information Forensics and Security}, vol.~19, pp. 2082--2097, 2024.

\bibitem{weissman2010actionCapacity}
T.~Weissman, ``Capacity of channels with action-dependent states,'' \emph{IEEE Transactions on Information Theory}, vol.~56, no.~11, pp. 5396--5411, 2010.

\bibitem{AhlswedeCaiWTCwithFeedback}
R.~Ahlswede and N.~Cai, ``Transmission, identification and common randomness capacities for wire-tape channels with secure feedback from the decoder,'' \emph{Electron. Notes Discrete Math.}, vol.~21, pp. 155--159, Aug. 2005.

\bibitem{AsafCohenWTCwithFeedback}
A.~Cohen and A.~Cohen, ``Wiretap channel with causal state information and secure rate-limited feedback,'' \emph{{IEEE} Trans. Commun.}, vol.~64, no.~3, pp. 1192--1203, Mar. 2016.

\bibitem{OurJSAITTutorial}
M.~Bloch, O.~G{\"u}nl{\"u}, A.~Yener, F.~Oggier, H.~V. Poor, L.~Sankar, and R.~F. Schaefer, ``An overview of information-theoretic security and privacy: {M}etrics, limits and applications,'' \emph{{IEEE} J. Sel. Areas Inf. Theory}, vol.~2, no.~1, pp. 5--22, 2021.

\bibitem{HanVinckWTCwithFeedback}
B.~Dai, A.~J.~H. Vinck, Y.~Luo, and Z.~Zhuang, ``Capacity region of non-degraded wiretap channel with noiseless feedback,'' in \emph{Proc. of {IEEE} Int. Symp. Inf. Theory}, Cambridge, MA, July 2012, pp. 244--248.

\bibitem{he-yener-fbsecrecy}
X.~{He} and A.~{Yener}, ``The role of feedback in two-way secure communications,'' \emph{{IEEE} Trans. Inf. Theory}, vol.~59, no.~12, pp. 8115--8130, Dec. 2013.

\bibitem{GermanWTCwithGeneralizedFeedback}
G.~Bassi, P.~Piantanida, and S.~Shamai, ``The wiretap channel with generalized feedback: {S}ecure communication and key generation,'' \emph{{IEEE} Trans. Inf. Theory}, vol.~65, no.~4, pp. 2213--2233, Apr. 2019.

\bibitem{YHKimWTCwithFeedback}
E.~Ardestanizadeh, M.~Franceschetti, T.~Javidi, and Y.-H. Kim, ``Wiretap channel with secure rate-limited feedback,'' \emph{{IEEE} Trans. Inf. Theory}, vol.~55, no.~12, pp. 5353--5361, Dec. 2009.

\bibitem{Tahmasbi2018}
M.~Tahmasbi, M.~R. Bloch, and A.~Yener, ``Learning an adversary's actions for secret communication,'' \emph{{IEEE} Trans. Inf. Theory}, vol.~66, no.~3, pp. 1607--1624, Mar. 2020.

\bibitem{zhang2019actionDependentfeedback}
H.~Zhang, L.~Yu, and B.~Dai, ``Feedback schemes for the action-dependent wiretap channel with noncausal state at the transmitter,'' \emph{Entropy}, vol.~21, no.~3, p. 278, 2019.

\bibitem{AhlswedeCsiz}
R.~Ahlswede and I.~Csisz{\'a}r, ``Common randomness in information theory and cryptography - {P}art {I}: Secret sharing,'' \emph{IEEE Trans. Inf. Theory}, vol.~39, no.~4, pp. 1121--1132, July 1993.

\bibitem{OSRBAmin}
M.~H. Yassaee, M.~R. Aref, and A.~Gohari, ``Achievability proof via output statistics of random binning,'' \emph{{IEEE} Trans. Inf. Theory}, vol.~60, no.~11, pp. 6760--6786, Nov. 2014.

\bibitem{RenesRenner}
J.~M. Renes and R.~Renner, ``Noisy channel coding via privacy amplification and information reconciliation,'' \emph{{IEEE} Trans. Inf. Theory}, vol.~57, no.~11, pp. 7377--7385, Nov. 2011.

\bibitem{RaviZivPartialSecrecyWTC}
J.~D.~D. Mutangana, R.~Tandon, Z.~Goldfeld, and S.~Shamai, ``Wiretap channel with latent variable secrecy,'' in \emph{Proc. of {IEEE} Int. Symp. Inf. Theory}, Melbourne, Australia, July 2021, pp. 837--842.

\bibitem{SW}
D.~Slepian and J.~Wolf, ``Noiseless coding of correlated information sources,'' \emph{{IEEE} Trans. Inf. Theory}, vol.~19, no.~4, pp. 471--480, July 1973.

\bibitem{Elgamalbook}
A.~E. Gamal and Y.-H. Kim, \emph{Network {I}nformation {T}heory}.\hskip 1em plus 0.5em minus 0.4em\relax Cambridge, {U.K.}: Cambridge {U}niversity {P}ress, 2011.

\bibitem{FMEbook}
A.~Schrijver, \emph{{T}heory of {L}inear and {I}nteger {P}rogramming}.\hskip 1em plus 0.5em minus 0.4em\relax Chichester, England: John Wiley \& Sons, June 1998.

\bibitem{CoverandThomas}
T.~M. Cover and J.~A. Thomas, \emph{Elements of {I}nformation {T}heory}, 2nd~ed.\hskip 1em plus 0.5em minus 0.4em\relax Hoboken, {NJ}: John {W}iley \& {S}ons, 2012.

\bibitem{CsiszarKornerbook2011}
I.~Csisz{\'a}r and J.~K{\"o}rner, \emph{Information {T}heory: {C}oding {T}heorems for {D}iscrete {M}emoryless {S}ystems}, 2nd~ed.\hskip 1em plus 0.5em minus 0.4em\relax Cambridge, {U.K}.: Cambridge {U}niversity {P}ress, 2011.

\bibitem{Chou2014d}
R.~A. Chou and M.~R. Bloch, ``Polar coding for the broadcast channel with confidential messages: {A} random binning analogy,'' \emph{{IEEE} Trans. Inf. Theory}, vol.~62, no.~5, pp. 2410--2429, May 2016.

\end{thebibliography}


\appendices

\section{Proof of Theorem \ref{theo:PSPOFPhysDegraded}}\label{app:thm1}

\begin{IEEEproof}
\textbf{Achievability:} We use the Output Statistics of Random Binning (OSRB) method \cite{OSRBAmin} to prove the achievability. We first define the operationally dual source coding problem to the problem of interest, the ISAC with transmitter actions channel coding problem. From there we define two protocols: Protocol A is coding scheme for the dual source coding problem and Protocol B is a randomized coding scheme for the original ISAC problem. 

Fix $p_{VAX}(v,a,x)$ such that there exist per-letter estimators $\Est_j(a,x,y_1,y_2)=~\hat{s}_j$ satisfying $\mathbb{E}[d_j(S_j,\widehat{S}_j))]\!\leq\!D_j + \epsilon_n$
  for $j=1,2$, where $\epsilon_n\!\geq0$ and $\epsilon_n\rightarrow0$ when $n\rightarrow\infty$. We now formally define Protocols A and B.

\textbf{Protocol A} \emph{(dual source coding problem)}: We generate the tuple of random variables $(V^n,A^n,X^n,Y_1^n,S_1^n,Y_2^n,S_2^n)$ i.i.d. according to \eqref{eq:jointprobPD}.

\emph{Random Binning:} The source encoder observing $v^n$, uniformly and independently assigns random bin indices $W_{\text{v}_1}\in[1:2^{nR_{\text{v}_1}}]$, $W_{\text{v}_2}\in[1:2^{nR_{\text{v}_2}}]$, $L_{\text{v}}\in[1:2^{n\overline{R}_{\text{v}}}]$, and $F_{\text{v}}\in[1:2^{n\widetilde{R}_{\text{v}}}]$. We also assign to $y_1^n$ a random bin index $L_{\text{y}_1}\in[1:2^{n\overline{R}_{\text{y}_1}}]$ and we set $\overline{R}_{\text{y}_1}=\overline{R}_{\text{v}}$. The legitimate receiver uses a Slepian-Wolf \cite{SW} decoder $P^{SW}(\hat{v}^n|y_1^n,s_1^n,f_{\text{v}})$ to recover the estimate $\hat{v}^n$ from $(y_1^n,s_1^n,f_{\text{v}})$. Since transmitter in the original problem has perfect output feedback, we assume the source encoder has access to $(y_1^n,y_2^n)$. Therefore, both the source encoder and the legitimate receiver can compute $L_{\text{y}_1}$. The eavesdropper observes $(y_2^n,s_2^n,f_{\text{v}})$.

The random pmf, denoted $P$, induced by the random binning in Protocol A is  
\begin{align}
    &P(v^n,a^n,x^n,y_1^n,s_1^n,y_2^n,s_2^n,w_{\text{v}_1},w_{\text{v}_2},l_{\text{v}},l_{\text{y}_1},f_{\text{v}},\hat{v}^n)\nonumber\\
    &=p(v^n,a^n,x^n,y_1^n,s_1^n,y_2^n,s_2^n)P(w_{\text{v}_1},w_{\text{v}_2},l_{\text{v}},f_{\text{v}}|v^n)\nonumber\\
    &\quad\quad\times P(l_{\text{y}_1}|y_1^n)P^{SW}(\hat{v}^n|y_1^n,s_1^n,f_{\text{v}})\nonumber\\
    &=P(w_{\text{v}_1},w_{\text{v}_2},l_{\text{v}},f_{\text{v}},v^n) p(a^n,x^n|v^n) \nonumber\\
    &\quad\quad\times P(y_1^n,s_1^n,y_2^n,s_2^n,l_{\text{y}_1}|a^n,x^n)P^{SW}(\hat{v}^n|y_1^n,s_1^n,f_{\text{v}})\nonumber\\
    &=P(w_{\text{v}_1},w_{\text{v}_2},l_{\text{v}},f_{\text{v}})P(v^n|w_{\text{v}_1},w_{\text{v}_2},l_{\text{v}},f_{\text{v}}) p(a^n,x^n|v^n)\nonumber\\
    &\quad\quad\times p(y_1^n,s_1^n,y_2^n,s_2^n|a^n,x^n)P(l_{\text{y}_1}|y_1^n)P^{SW}(\hat{v}^n|y_1^n,s_1^n,f_{\text{v}}). \label{eq:ProtocolAinitialdistribution}
\end{align}

\textbf{Protocol B} \emph{(channel coding problem assisted with shared randomness)}: Assume the existence and public dissemination of the shared randomness $F_{\text{v}}$ to all parties, the transmitter, legitimate receiver, and the eavesdropper, where $F_{\text{v}}$ distributed uniformly over $[1:2^{n\widetilde{R}_{\text{v}}}]$. We also assume the existence of a secret key $K$, uniformly selected from $[1:2^{n\overline{R}_{\text{v}}}]$, which is securely shared between the transmitter and legitimate receiver. We will later remove $K$ using a block-Markov coding scheme, in which we will replace $K$ with $L_{\text{y}_1}$. Conceptually, we define the message $M=(M_1,M_2)$ choosing the public message $M_1=W_{\text{v}_1}$, i.e. the message with no security requirement, and the private message $M_2=(W_{\text{v}_2},L)$, i.e. the message that should be secured from the eavesdropper. Note that $L$ is not a random bin index.

The coding scheme proceeds as follows: the transmitter selects two messages $m_1\in\text{Unif}[1:2^{nR_{\text{v}_1}}]$ and $m_2\in\text{Unif}[1:2^{n(R_{\text{v}_2}+\overline{R}_{\text{v}})}]$ independent of each other and $F_{\text{v}}$. The messages can be represented as $m_1=w_{\text{v}_1}$ and $m_2=(w_{\text{v}_2},l)$, where $l\in[1:2^{n\overline{R}_{\text{v}}}]$. The transmitter obtains $l_{\text{v}}$ by securing $l$ via the one-time padding operation using $K$, i.e.,  $l_{\text{v}}=l\oplus K$. The transmitter generates $v^n$ according to  $P(v^n|w_{\text{v}_1},w_{\text{v}_2},l_{\text{v}},F_{\text{v}})$ as in \eqref{eq:ProtocolAinitialdistribution}. After $n$-channel uses, the transmitter, legitimate receiver, and eavesdropper observe $(v^n,a^n,x^n,y_1^n,y_2^n,f_{\text{v}})$, $(y_1^n,s_1^n,f_{\text{v}})$, and $(y_2^n,s_2^n,f_{\text{v}})$, respectively. From here both the transmitter and legitimate receiver can generate $l_{\text{y}_1}$ according to $P(l_{\text{y}_1}|y_1^n)$ from \eqref{eq:ProtocolAinitialdistribution}. The legitimate receiver will use the same Slepian-Wolf decoder from Protocol A, $P^{SW}(\hat{v}^n|y_1^n,s_1^n,f_{\text{v}})$, to find its estimate $\hat{v}^n$.

The distribution induced by Protocol B, denoted $\widehat{P}$, is
\begin{align}
    &\widehat{P}(v^n,a^n,x^n,y_1^n,s_1^n,y_2^n,s_2^n,w_{\text{v}_1},w_{\text{v}_2},l_{\text{v}},l_{\text{y}_1},f_{\text{v}},\hat{v}^n)\nonumber\\
    &=\text{Unif}[1:2^{n\widetilde{R}_{\text{v}}}]\cdot \text{Unif}[1:2^{nR_{\text{v}_1}}] \cdot \text{Unif}[1:2^{n(R_{\text{v}_2}+\overline{R}_{\text{v}})}]\nonumber\\
    &\quad\quad\times P(v^n|w_{\text{v}_1},w_{\text{v}_2},l_{\text{v}},f_{\text{v}}) p(a^n,x^n|v^n)\nonumber\\
    &\quad\quad\times p(y_1^n,s_1^n,y_2^n,s_2^n|a^n,x^n)P(l_{\text{y}_1}|y_1^n)P^{SW}(\hat{v}^n|y_1^n,s_1^n,f_{\text{v}}). \label{eq:ProtocolBinitialdistribution}
\end{align}

The distributions induced by Protocols A and B are approximately the same when the random binning indices $(W_{\text{v}_1},W_{\text{v}_2},L_{\text{v}},F_{\text{v}})$ are almost mutually independent and uniformly distributed. More specifically, following \cite[Theorem 1]{OSRBAmin}, as $n\rightarrow\infty$ the total variational distance between $P$ and $\hat{P}$ goes to zero when
\begin{align}
    R_{\text{v}_1} + R_{\text{v}_2} + \overline{R}_{\text{v}} + \widetilde{R}_{\text{v}} < H(V). \label{eq:UniformityIndepndenceOfAllBinIndices}
\end{align}

Now we find the conditions that guarantee reliability and security for Protocol A.

\emph{Reliability}: The rate condition that guarantees reliability is exactly that for which the Slepian-Wolf can successfully recover $v^n$ from $(y_1^n,s_1^n,f_{\text{v}})$. By \cite[Lemma 1]{OSRBAmin}, imposing the constraint
\begin{align}
    \widetilde{R}_{\text{v}} > H(V|Y_1,S_1) \label{eq:AchReliabilityratecondition}
\end{align}
implies that in the expectation over the random binning the total variational distance between $P^{SW}(\hat{v}^n|y_1^n,s_1^n,f_{\text{v}})$ and $\mathbbm{1}\{v^n = \hat{v}^n\}$ goes to zero as $n\rightarrow\infty$, giving reliability for Protocol A. Reliability for Protocol B follows from triangle inequality and a combination of the reliability for protocol A and the vanishing expected variational distance between \eqref{eq:ProtocolAinitialdistribution} and \eqref{eq:ProtocolBinitialdistribution}.

\emph{Security}: In order to show that Protocol B is secure, we find the rate conditions that show $L_{\text{y}_1}$ and $W_{\text{v}_2}$ are secure. To do this, we show that the random binning indices are independent of the eavesdroppers observations and uniformly distributed. We later show that $L$ is secure through application of the secret key $K$.

Using \cite[Theorem 1]{OSRBAmin}, $L_{\text{y}_1}$ becomes almost independent of $(V^n,Y_2^n,S_2^n)$ and uniformly distributed if 
\begin{align}
    \overline{R}_{\text{y}_1}=\overline{R}_{\text{v}} < H(Y_1|V,Y_2,S_2). \label{eq:SecretKeySecurityRateCondition}
\end{align}
More specifically, \eqref{eq:SecretKeySecurityRateCondition} in the expectation over the random binning, the total variational distance between
$P(L_{\text{v}},V^n,Y_2^n,S_2^n) \label{eq:UnsecuredKeyDistribution}$
and
$\text{Unif}[1:2^{n\overbar{R}_{\text{y}_1}}]\cdot p(V^n,Y_2^n,S_2^n)$
goes to zero as $n\rightarrow\infty$. 

A similar application of \cite[Theorem 1]{OSRBAmin} shows indices $F_\text{v}$ and $W_{\text{v}_2}$ become almost independent of $(Y_2^n,S_2^n)$ and uniformly distributed if 
\begin{align}
    \widetilde{R}_{\text{v}} + R_{\text{v}_2} < H(V|Y_2,S_2) \label{eq:W2secureratecondition}
\end{align}
i.e., in the expectation over the random binning, the total variational distance between
$P(F_{\text{v}},W_{\text{v}_2},Y_2^n,S_2^n)$
and
$\text{Unif}[1:2^{n\widetilde{R}_{\text{v}}}]\cdot\text{Unif}[1:2^{nR_{\text{v}_2}}]\cdot p(Y_2^n,S_2^n)$
goes to zero as $n\rightarrow\infty$.

Using the properties of variational distance between distributions found in \cite[Lemma 4]{OSRBAmin}, it follows that $W_{\text{v}_2}$ and $L_{\text{y}_1}$ are also secure in Protocol B. The uniformity and independence of $K$ with all other random variables in addition to the uniformity of $L$ means the one-time padding secures $L$ from the eavesdropper, i.e. makes $L$ independent of $(Y_2^n,S_2^n,F_{\text{v}})$.

Performing Fourier-Motzkin elimination \cite{FMEbook} on \eqref{eq:UniformityIndepndenceOfAllBinIndices}-\eqref{eq:W2secureratecondition} to remove $\widetilde{R}_{\text{v}}$ gives
\begin{align}
    & R_{\text{v}_1}  < I(V;Y_1,S_1) - R_{\text v_2} - \overbar{R}_{\text{v}},\\
    & R_{\text{v}_2}  < \big[H(V|Y_2,S_2)-H(V|Y_1,S_1)\big]^+, \label{eq:WTRate}\\
    &\overbar{R}_{\text{v}}  < H(Y_1|Y_2,S_2,V)
\end{align}
where $[b]^+=\max\big\{0,b\big\}$, and \eqref{eq:WTRate} follows since all rates must be non-negative. For any $\epsilon>0$ the rate
\begin{align}
    R_1 = R_{\text{v}_1} & = I(V;Y_1,S_1) - \epsilon  \label{eq:R1BoundStep}
\end{align}
is achievable. For any $R_1$ less than or equal to \eqref{eq:R1BoundStep}, the rate
\begin{align}
    R_2=R_{\text{v}}+\overbar{R}_{\text{v}} = \min\big\{R_{\text{sec}},I(V;Y_1,S_1)-R_1\big\}-2\epsilon,\nonumber
\end{align}
where 
\begin{align}
   R_{\text{sec}} = \big[H(V|Y_2,S_2) - H(V|Y_1,S_1)\big]^+ +H(Y_1|Y_2,S_2,V)  \label{eq:R2BoundStep}
\end{align}
is achievable.

The channel with no degradedness assumptions conforms to the Markov chain in \eqref{eq:channelMC}, which, when combined with \eqref{eq:physicaldegradedcond} and \eqref{eq:channelMC} provides the Markov chain \eqref{eq:pdmc}.

We can now perform the following simplifications
\begin{align}
    &R_{\text{sec}}  = \big[H(V|Y_2,S_2) - H(V|Y_1,S_1)\big]^+ +H(Y_1|Y_2,S_2,V)\nonumber\\
    & \overset{(a)}{=} H(V|Y_2,S_2) - H(V|Y_1,S_1) +H(Y_1|Y_2,S_2,V)\nonumber\\
    & \overset{(b)}{=} H(V,Y_1|Y_2,S_2) - H(V|Y_1,S_1,Y_2,S_2)\nonumber\\
    & = H(Y_1|Y_2,S_2) + I(V;S_1|Y_1,Y_2,S_2)\nonumber\\
    & = H(Y_1|Y_2,S_2) + H(S_1|Y_1,Y_2,S_2) - H(S_1|Y_1,Y_2,S_2,V)\nonumber\\
    & = H(Y_1,S_1|Y_2,S_2) - H(S_1|Y_1,Y_2,S_2,V)
\end{align}
where $(a)$ follows from the data processing inequality \cite{CoverandThomas} and \eqref{eq:pdmc}, and $(b)$ follows from \eqref{eq:pdmc}. Thus, we achieve \eqref{eq:achPD1} and \eqref{eq:achPD2}.

The distortion constraints follow since all $(v^n,x^n,y_1^n,y_2^n,s_1^n,s_2^n)$ tuples are in the jointly typical set w. h. p. Using the law of total expectation to bounded distortion metrics and the typical average lemma \cite[pp.~26]{Elgamalbook}, we see that the distortion constraints \eqref{eq:distortion_cons} are satisfied. The sufficiency of the deterministic estimators in \eqref{eq:deterministicest} follows from \cite[Lemma 1]{MariMichelleGJournal}, where $(S^n,(X^n,Z^n),\hat{S}^n)$ is replaced with $(S_j^n,(X^n,A^n,Y_1^n,Y_2^n),\hat{S}_j^n)$ for $j=1,2$ and noting that $(X^{n\setminus i},A^{n\setminus i},Y^{n\setminus i}_1,Y^{n\setminus i}_2,\hat{S}_{j,i})-(X_i,A_i,Y_{1,i},Y_{2,i})-S_{j,i}$ forms a Markov chain.

Now we need to derandomize Protocol B and remove the dependence on a secret key. The existence of a specific realization $f$ of $F$ such that the reliability and secrecy properties of B still holds follow from the standard method in \cite{OSRBAmin}.

Finally, we remove the secret key from Protocol~B by chaining over multiple blocks as in \cite{Chou2014d} and \cite{gunlu2023secureISAC}. We will use a block-Markov coding scheme consisting of $B\!\geq\!2$ blocks, each with $n$ channel uses. For the discussion of the block-Markov coding scheme alone, we will denote the random variables corresponding to the transmissions in block $b$ with a superscript $b$ and a sequence of random variables from blocks $i$ to $j$, with $0\leq i\leq j \leq B$, denoted with a superscript $i:j$, e.g., the random bin index for $W_{\text{v}_1}$ block $b$ is $W_{\text{v}_1}^b$. We let $M=M^{1:B}=(M_1^{1:B},M_2^{1:B})$. Noting that in each block $b$, a secret key $L_{\text{y}_1}^b$ is generated known to the transmitter and legitimate receiver after the completion of block $b$. We set $M_2^1=\emptyset$, i.e., no secure message is sent during the first block, and $L_{\text{y}_1}^{b-1}$ is used as $K$ from Protocol B during block $b\geq2$.

Since the distortion constraints are satisfied for a single block, they will be satisfied when aggregated across all blocks. The rate change for $R_2$ is negligible for $B$ large, while $R_1$ remains the same. We finally need to show that the reliability and secrecy performance is unaffected by the block-Markov coding scheme. The asymptotic reliability performance follows from a union bound on 
\begin{align}
    \lim\limits_{n\rightarrow\infty}P\bigg[\bigg\{\widehat{M}_1^{1:B}\neq M_1^{1:B} \text{ or }\widehat{M}_2^{1:B}\neq M_2^{1:B}\bigg\}\bigg].
\end{align}

The single block security analysis shows that leakage between $M_2$ and the eavesdropper's observations within a single block is negligible, but, in order to show that the secrecy performance is unaffected by the block-Markov coding scheme we need to show that the leakage between the secure message, $M_2^{1:B}=(W_{\text{v}_2}^{1:B},L^{1:B})$, and the eavesdropper's observations over all blocks, $(\mathbf{Y}_2^{1:B},\mathbf{S}_2^{1:B})$, where the bold faced random variables are $n$-letter random variables, is negligible, i.e., $I(W_{\text{v}_2}^{1:B},L^{1:B};\mathbf{Y}_2^{1:B},\mathbf{S}_2^{1:B})$ vanishes asymptotically. We make slight modifications to the approach in \cite{gunlu2023secureISAC} to show that security holds across all blocks.

For convenience, we denote $W^b=W_{\text{v}_2}^b$, $K^b=L_{\text{y}_1}^b$, $L_{\text{v}}^b= L^b\oplus K^b$, and $\mathbf{Z}^{b}=(\mathbf{Y}_2^{b},\mathbf{S}_2^{b})$. We have
\begin{align}
    &I(W^{1:B},L^{1:B};\mathbf{Z}^{B})\nonumber\\
    & =  \sum_{b=1}^{B-1}\left( I(W^{1:B},L^{1:B};\mathbf{Z}^{1:b+1}) -  I((W^{1:B},L^{1:B};\mathbf{Z}^{1:b})\right) \nonumber\\
    &\qquad\qquad + I(W^{1:B},L^{1:B};\mathbf{Z}^{1})\nonumber\\
    &\overset{(a)}{=}  \sum_{b=1}^{B-1}\left( I(W^{1:B},L^{1:B};\mathbf{Z}^{1:b+1}) -  I((W^{1:B},L^{1:B};\mathbf{Z}^{1:b})\right) \label{eq:InitialSumBlockSecrecy}
\end{align}
where $(a)$ follows since $\mathbf{Z}^1$ is independent of future messages $(W^{2:B},L^{2:B})$ and no secure message is transmitted in the first block. Without loss of generality, we consider the term in the sum corresponding to block $b$, we see
\begin{align}
    &I(W^{1:B},L^{1:B};\mathbf{Z}^{1:b+1}) -  I(W^{1:B},L^{1:B};\mathbf{Z}^{1:b})\nonumber\\
    &= I(W^{1:B},L^{1:B};\mathbf{Z}^{b+1}|\mathbf{Z}^{1:b})\nonumber\\
    &= I(W^{1:b+1},L^{1:b+1};\mathbf{Z}^{b+1}|\mathbf{Z}^{1:b})\nonumber\\
    &\qquad\qquad +I(W^{b+2:B},L^{b+2:B};\mathbf{Z}^{b+1}|\mathbf{Z}^{1:b},W^{1:b+1},L^{1:b+1})\nonumber\\
    &\leq I(W^{1:b+1},L^{1:b+1},\mathbf{Z}^{1:b};\mathbf{Z}^{b+1})\nonumber\\
    &\qquad\qquad +I(W^{b+2:B},L^{b+2:B};\mathbf{Z}^{1:b+1},W^{1:b+1},L^{1:b+1})\nonumber\\
    &\stackrel{(a)}{=} I(W^{1:b+1},L^{1:b+1},\mathbf{Z}^{1:b};\mathbf{Z}^{b+1})\nonumber\\
    &=I(W^{b+1},L^{b+1};\mathbf{Z}^{b+1})\nonumber\\
    &\qquad\qquad  +I(W^{1:b},L^{1:b},\mathbf{Z}^{1:b};\mathbf{Z}^{b+1}|W^{b+1},L^{b+1})\nonumber\\
    &\stackrel{(b)}{=}I(W^{b+1},L^{b+1};\mathbf{Z}^{b+1})\nonumber\\
    &\qquad\qquad +I(W^{1:b},L^{1:b},\mathbf{Z}^{1:b};\mathbf{Z}^{b+1},W^{b+1},L^{b+1})\nonumber\\
    &\stackrel{(c)}{\leq}I(W^{b+1},L^{b+1};\mathbf{Z}^{b+1})\nonumber\\
    &\qquad\qquad +I(W^{1:b},L^{1:b},\mathbf{Z}^{1:b};\mathbf{Z}^{b+1},W^{b+1},L^{b+1},K^b)\nonumber\\
    &=I(W^{b+1},L^{b+1};\mathbf{Z}^{b+1})+I(W^{1:b},L^{1:b},\mathbf{Z}^{1:b};K^b)\nonumber\\
    &\qquad\qquad +I(W^{1:b},L^{1:b},\mathbf{Z}^{1:b};\mathbf{Z}^{b+1},W^{b+1},L^{b+1}|K^b)\nonumber\\
    &\stackrel{(d)}{=}I(W^{b+1},L^{b+1};\mathbf{Z}^{b+1})+I(W^{1:b},L^{1:b},\mathbf{Z}^{1:b};K^b)\nonumber\\
    &\stackrel{(e)}{\leq}I(W^{b+1},L^{b+1};\mathbf{Z}^{b+1})\!+\!I(W^{1:b},L^{1:b},\mathbf{Z}^{1:b},K^{b-1};K^b)\nonumber\\
    &\stackrel{(f)}{=}\!I(W^{b+1}\!,L^{b+1};\mathbf{Z}^{b+1})\!+\!I(W^{b},L^{b},\mathbf{Z}^{b},K^{b-1};K^b) \label{eq:SingleTermBlockSeecrecy}
\end{align}
where $(a)$ and $(b)$ follow since future messages are independent of past messages and observations; $(c)$ follows by adding a non-negative term in order to introduce $K^b$ to break dependence across blocks; $(d)$ follows since 
\begin{align}
    (W^{1:b},L^{1:b},\mathbf{Z}^{1:b}) - K^b - (\mathbf{Z}^{b+1},W^{b+1},L^{b+1})
\end{align} 
forms a Markov chain; $(e)$ follows by adding $K^{b-1}$ in again to break dependence across blocks, and $(f)$ follows since we have the Markov chain 
\begin{align}
    (W^{1:b-1},L^{1:b-1},\mathbf{Z}^{1:b-1})-(W^{b},L^{b},\mathbf{Z}^{b},K^{b-1})-K_b.
\end{align}

Combining \eqref{eq:SingleTermBlockSeecrecy} with \eqref{eq:InitialSumBlockSecrecy} gives
\begin{align}
    &I(W^{1:B},L^{1:B};\mathbf{Z}^{1:B}) \nonumber\\
    &\leq \sum\limits_{b=1}^{B-1}\bigg(I\big(W^{b+1},L^{b+1};\mathbf{Z}^{b+1}\big)+ I\big(W^b,L^b,\mathbf{Z}^b,K^{b-1};K^b\big)\bigg).
\end{align}
Now we show that the two quantities $I\big(W^{b+1},L^{b+1};\mathbf{Z}^{b+1}\big)$ and $I\big(W^b,L^b,\mathbf{Z}^b,K^{b-1};K^b\big)$ asymptotically vanish across all blocks.
The first term can be written as
\begin{align}
    &I\big(W^{b+1},L^{b+1};\mathbf{Z}^{b+1}\big) \nonumber\\
    &= I\big(W^{b+1};\mathbf{Z}^{b+1}\big) + I\big(L^{b+1};\mathbf{Z}^{b+1}|W^{b+1}\big) \nonumber\\
    &\overset{(a)}{=} I\big(W^{b+1};\mathbf{Z}^{b+1}\big) + I\big(L^{b+1};\mathbf{Z}^{b+1},W^{b+1}\big) \nonumber\\
    &\overset{(b)}{=} I\big(W^{b+1};\mathbf{Z}^{b+1}\big) + I\big(L^{b+1};L_{\text{v}}^b\big) \nonumber\\
    &\overset{(c)}{=} I\big(W^{b+1};\mathbf{Z}^{b+1}\big) + H\big(L_{\text{v}}^b\big) - H\big(K^b\big) \nonumber\\
    &\overset{(d)}{\leq} I\big(W^{b+1};\mathbf{Z}^{b+1}\big) + n\overbar{R}_{\text{v}} - H\big(K^b\big)
\end{align}
where $(a)$ follows since $W^{b+1}$ and $L^{b+1}$ are independent, $(b)$ follows by the data processing inequality and $L^{b+1}-L_{\text{v}}^b-(\mathbf{Z}^{b+1},W^{b+1})$, $(c)$ follows since $H(L_{\text{v}}^b|L^b)=H(K^b)$, and $(d)$ follows from the rate condition on $L_{\text{v}}^b$. The condition \eqref{eq:W2secureratecondition} guarantees that $I(W^{b+1};\mathbf{Z}^{b+1})$ vanishes across all blocks, while \eqref{eq:SecretKeySecurityRateCondition} makes $K^b$ uniform, causing $n\overbar{R}_{\text{v}} - H\big(K^b\big)$ to vanish across all blocks. The independence of the secret key with other random variables $K^b=L_{\text{y}_1}$ in a single block, see \eqref{eq:SecretKeySecurityRateCondition}, implies that $I\big(W^b,L^b,\mathbf{Z}^b,K^{b-1};K^b\big)$ vanishes across all blocks as $n\rightarrow\infty$.

\vspace{.5em}
\noindent\textbf{Converse:}
For the converse proof, assume that for some $\delta_n>0$, with $\delta_n\rightarrow 0$ as $n\rightarrow\infty$, there exist an action encoder, channel encoder, decoder, and estimators such that \eqref{eq:rates_cons}-\eqref{eq:distortion_cons} are satisfied for $(R_1,R_2,D_1,D_2)$. We define 
\begin{align}
    V_{i}\triangleq (M_1,M_2,Y^{i-1}_{1},S_1^{i-1},Y^{i-1}_{2},S_2^{i-1})\label{eq:DefinitionofVi}
\end{align} 
such that $V_i-(A_i,X_i)-(Y_{1,i},Y_{2,i},S_{1,i},S_{2,i})$ forms a Markov chain for all $i\in[1:~n]$.

Using Fano's inequality \cite{CoverandThomas} we obtain
\begin{align}
    H(M;Y_1^n,S_1^n)&\leq n\epsilon_n\label{eq:fano}
\end{align}
where $n\epsilon_n=H_b(\delta_n) + \delta_n(R_1+R_2)n$. Note that $\epsilon_n\rightarrow0$ as $\delta_n\rightarrow0$.

\emph{Bound on $R_1$}: We have 
\begin{align}
  & nR_1  \overset{(a)}{\leq} I(M_1;Y_1^n,S_1^n) + n\epsilon_n \nonumber\\
       & = \sum\limits_{i=1}^n\big[H\big(Y_{1,i},S_{1,i}|Y_1^{i-1},S_1^{i-1}\big)\nonumber\\
       & \qquad\qquad -H\big(Y_{1,i},S_{1,i}|M_1,Y_1^{i-1},S_1^{i-1}\big) +\epsilon_n\big]\nonumber\\
       & \leq \sum\limits_{i=1}^n\big[H\big(Y_{1,i},S_{1,i}\big)\nonumber\\
       & \qquad\qquad -H\big(Y_{1,i},S_{1,i}|M,Y_1^{i-1},S_1^{i-1},Y_2^{i-1},S_2^{i-1}\big) +\epsilon_n\big]\nonumber\\
       & \overset{(b)}{=} \sum\limits_{i=1}^n\big[I\big(V_i;Y_{1,i},S_{1,i}\big) + \epsilon_n\big] \label{eq:R1BoundConvApp}
\end{align}
where $(a)$ follows from Fano's inequality \eqref{eq:fano} and $(b)$ follows from the definition of $V_i$ \eqref{eq:DefinitionofVi}.

\emph{Bound on $R_1+R_2$}: Similar to the bound on $R_1$, we have
\begin{align}
  &n(R_1+R_2) \overset{(a)}{\leq} I(M;Y_1^n,S_1^n) + n\epsilon_n \nonumber\\
    & = \sum\limits_{i=1}^n\big[H\big(Y_{1,i},S_{1,i}|Y_1^{i-1},S_1^{i-1}\big)\nonumber\\
    & \qquad\qquad -H\big(Y_{1,i},S_{1,i}|M,Y_1^{i-1},S_1^{i-1}\big) +\epsilon_n\big]\nonumber\\
    & \leq \sum\limits_{i=1}^n\big[H\big(Y_{1,i},S_{1,i}\big)\nonumber\\
    & \qquad\qquad -H\big(Y_{1,i},S_{1,i}|M,Y_1^{i-1},S_1^{i-1},Y_2^{i-1},S_2^{i-1}\big) +\epsilon_n\big]\nonumber\\
    & \overset{(b)}{=} \sum\limits_{i=1}^n\big[I\big(V_i;Y_{1,i},S_{1,i}\big) + \epsilon_n\big] \label{eq:R12BoundConv}
\end{align}
where $(a)$ follows from Fano's inequality \eqref{eq:fano} and $(b)$ follows from the definition of $V_i$ \eqref{eq:DefinitionofVi}.

\emph{Bound on $R_2$}: We obtain
\begin{align}
  & nR_2 \overset{(a)}{\leq} I(M_2;Y_1^n,S_1^n,Y_2^n,S_2^n) + n\epsilon_n \nonumber\\
  & = H(Y_1^n,S_1^n|Y_2^n,S_2^n) + I(M_2;Y_2^n,S_2^n) \nonumber\\
  &\quad\quad- H(Y_1^n,S_1^n|M_2,Y_2^n,S_2^n) + n\epsilon_n \nonumber\\
  & \overset{(b)}{\leq} H(Y_1^n,S_1^n|Y_2^n,S_2^n) + \delta_n \nonumber\\
  &\quad\quad- H(S_1^n|M,Y_1^n,Y_2^n,S_2^n) +n\epsilon_n \nonumber\nonumber\\
  & \leq \sum\limits_{i=1}^n\big[H(Y_{1,i},S_{1,i}|Y_{2,i},S_{2,i})\nonumber\\
  &\qquad\qquad- H(S_{1,i}|M,Y_1^n,S_1^{i-1},Y_2^n,S_2^n)\big] + n\epsilon_n + \delta_n\nonumber\\
  & \overset{(c)}{=} \sum\limits_{i=1}^n\big[H(Y_{1,i},S_{1,i}|Y_{2,i},S_{2,i})\nonumber\\
  &\qquad\qquad- H(S_{1,i}|M,Y_1^i,S_1^{i-1},Y_2^i,S_2^i)\big] + n\epsilon_n + \delta_n\nonumber\\
  & \overset{(d)}{=} \sum\limits_{i=1}^n\big[H(Y_{1,i},S_{1,i}|Y_{2,i},S_{2,i})\nonumber\\
  &\qquad\qquad- H(S_{1,i}|M,Y_{1,i},Y_{2,i},S_{2,i},V_i)\big] + n\epsilon_n + \delta_n\label{eq:BoundR2}
\end{align}
where $(a)$ follows from Fano's inequality \eqref{eq:fano}, $(b)$ follows since the secrecy constraint \eqref{eq:secrecyleakage_cons} is satisfied, and $(c)$ is a consequence of the Markov chain 
\begin{align*}
(Y_{1,i+1}^n,Y_{2,i+1}^n,S_{2,i+1}^n)-(M,Y_1^i,S_1^{i-1},Y_2^i,S_2^i)-S_{1,i}
\end{align*}
and $(d)$ follows from definition of $V_i$, see \eqref{eq:DefinitionofVi}.

The deterministic estimators, for j=1,2, follow from
\begin{align}
  D_j+\delta_n \geq \mathbb{E}\big[d_j(S_j^n,\widehat{S_j^n})\big]=\frac{1}{n}\sum\limits_{i=1}^n\mathbb{E}\big[d_j(S_{j,i},\widehat{S}_{j,i})\big].
\end{align}

Next, we introduce a time-sharing random variable $Q$ distributed uniformly on $[1:n]$ which is independent of all other random variables. This allows us to represent the bounds on $R_1$ \eqref{eq:R1BoundConvApp} and $R_1+R_2$ \eqref{eq:R12BoundConv} as
\begin{align}
    &R_1 \leq\sum\limits_{i=1}^n\big[I\big(V_Q;Y_{1,Q},S_{1,Q}|Q=i\big) + \epsilon_n\big] \nonumber\\
    & = I\big(V_Q;Y_{1,Q},S_{1,Q}|Q\big) + \epsilon_n \nonumber\\
    & = I\big(V_Q,Q;Y_{1,Q},S_{1,Q}\big) - I\big(Q;Y_{1,Q},S_{1,Q}\big) + \epsilon_n \nonumber\\
    & \leq I\big(V_Q,Q;Y_{1,Q},S_{1,Q}\big) + \epsilon_n
\end{align}
and
\begin{align}
    R_1+R_2 \leq I\big(V_Q,Q;Y_{1,Q},S_{1,Q}\big) + \epsilon_n
\end{align}
and the bound on $R_2$ given in \eqref{eq:BoundR2} as
\begin{align}
    &R_2 \leq \sum\limits_{i=1}^n\big[H(Y_{1,i},S_{1,i}|Y_{2,i},S_{2,i}) \!+\!\epsilon_n\!+\!\frac{1}{n}\delta_n\nonumber\\
    &\quad\qquad\qquad- H(S_{1,i}|M,Y_{1,i},Y_{2,i},S_{2,i},V_i)\big]\nonumber\\
    &= \sum\limits_{i=1}^n\big[H(Y_{1,Q},S_{1,Q}|Y_{2,Q},S_{2,Q},Q=i) \!+\!\epsilon_n\!+\!\frac{1}{n}\delta_n\nonumber\\
    &\qquad\qquad- H(S_{1,Q}|M,Y_{1,Q},Y_{2,Q},S_{2,Q},V_Q,Q=i)\big]\nonumber\\
    &= H(Y_{1,Q},S_{1,Q}|Y_{2,Q},S_{2,Q},Q)\nonumber\\
    &\qquad- H(S_{1,i}|M,Y_{1,Q},Y_{2,Q},S_{2,Q},V_Q,Q) + n\epsilon_n + \delta_n\nonumber\\
    &\leq H(Y_{1,Q},S_{1,Q}|Y_{2,Q},S_{2,Q})\nonumber\\
    &\qquad- H(S_{1,i}|M,Y_{1,Q},Y_{2,Q},S_{2,Q},V_Q,Q) + n\epsilon_n\! +\! \delta_n.
\end{align}

Defining $X=X_Q$, $A=A_Q$, $Y_1=Y_{1,Q}$, $S_1=S_{1,Q}$, $Y_1=Y_{1,Q}$, $S_2=S_{2,Q}$, and $V=(V_Q,Q)$ so that $V-(X,A)-(Y_1,S_1,Y_2,S_2)$ forms a Markov chain and letting $\delta_n\rightarrow0$ gives the conditions in the statement of the theorem, \eqref{eq:achPD1}-\eqref{eq:achdistortion1and2}.

The proof of the cardinality bound for $V$ follows from the support lemma \cite[Lemma 15.4]{CsiszarKornerbook2011}.
\end{IEEEproof}


\section{Proof of Theorem~\ref{theo:PSPOFRevPhysDegraded}}\label{app:thm2}
\begin{IEEEproof}
The proof of Theorem~\ref{theo:PSPOFRevPhysDegraded} follows similarly to that of Theorem~\ref{theo:PSPOFPhysDegraded} in Appendix~\ref{app:thm1}, we highlight the modifications below.

\noindent\textbf{Achievability:} The creation of Protocols A and B follow as in Appendix~\ref{app:thm1} up to the recovery of the rate conditions \eqref{eq:R1BoundStep} and \eqref{eq:R2BoundStep}. Combining the definition of reverse-physical degradation in \eqref{eq:reversephysicallydegradedcond} with the channel Markov chain \eqref{eq:channelMC} gives the Markov chain \eqref{eq:rpdmc}, which we can use to perform the following simplification on $R_{sec}$ from \eqref{eq:R2BoundStep}
\begin{align}
    &R_{\text{sec}} = \big[H(V|Y_2,S_2) - H(V|Y_1,S_1)\big]^+ +H(Y_1|Y_2,S_2,V)\nonumber\\
    & \overset{(a)}{=} H(Y_1|Y_2,S_2,V)\nonumber\\
    & \overset{(b)}{=} H(Y_1|Y_2,S_2) 
\end{align}
where $(a)$ follows since an application of \eqref{eq:rpdmc} and the data processing inequality implies that $H(V|Y_2,S_2) \leq H(V|Y_1,S_1)$, and $(b)$ follows from \eqref{eq:rpdmc}. We thus have the rate conditions of the theorem, \eqref{eq:achPD1}, \eqref{eq:achdistortion1and2}, and \eqref{eq:achPDR2}. The distortion constraints, viability of per-letter deterministic estimators, derandomization, and analysis of the block-Markov coding scheme follow from the same argument in Appendix~\ref{app:thm1}.

\noindent\textbf{Converse:}
Noting that the converse proof in Appendix~\ref{app:thm1} does not use degradation, thus, the rate constraint on $R_1$ \eqref{eq:achPD1} and distortion constraints \eqref{eq:achdistortion1and2} holds here as well. We can simplify constraint on $R_2'$, see \eqref{eq:R2primedef}, in \eqref{eq:achPD2} as
\begin{align}
    &R_{2}^{\prime}\leq H(Y_1,S_1|Y_2,S_2)-H(S_1|Y_1,Y_2,S_2,V)\nonumber\\
    & = H(Y_1,S_1|Y_2,S_2)\!-\!H(S_1,Y_1|Y_2,S_2,V)\!+\!H(Y_1|Y_2,S_2,V)\nonumber\\
    & \overset{(a)}{=} H(Y_1,S_1|Y_2,S_2)\!-\!H(S_1,Y_1|Y_2,S_2)\!+\!H(Y_1|Y_2,S_2)\nonumber\\
    & = H(Y_1|Y_2,S_2)
  \end{align}
where $(a)$ follows from the reverse-physical degradation, see \eqref{eq:rpdmc}. This gives
\begin{align}
    R_2\leq\min\big\{H(Y_1|Y_2,S_2),I(V;Y_1,S_1)-R_1\big\}
\end{align}
giving the converse.
\end{IEEEproof}


\section{Proof of Theorem~\ref{theo:PDfullsecrecy}}\label{app:thm3}
\begin{IEEEproof}
\textbf{Achievability:} The achievability loosely follows by removing $V^n$ and using $(X^n,A^n)$ in its place in the proof of Theorem~\ref{theo:PSPOFPhysDegraded}.

We fix $P_{AX}$ and generate a tuple of random variables i.i.d. according to \eqref{eq:physicaldegradedcond}. Protocols A and B are defined as in Appendix~\ref{app:thm1} with the following modifications. The random binning is performed on $(a^n,x^n)$ instead of $v^n$, the random binning index $W_{\text{v}_1}$ is removed, $W_{\text{v}_2}$ is replaced with $W_{\text{ax}}\in\textnormal{Unif}[1:2^{nR_{\text{ax}}}]$,  $F_{\text{v}}$ is replace with $F_{\text{ax}}\in\textnormal{Unif}[1:2^{n\widetilde{R}_{\text{ax}}}]$, and $L_\text{v}$ is replaced by $L_{\text{ax}}\in\textnormal{Unif}[1:2^{n\overbar{R}_{\text{ax}}}]$. We continue to use $L$ and $L_{\text{y}_1}$ defined therein. We define $M=(W_{\text{ax}},L)$, the entirety of which should be kept secret from the eavesdropper.

The distributions induced by Protocols A and B are approximately the same when 
\begin{align}
    R_{\text{ax}} + \overbar{R}_{\text{ax}} + \widetilde{R}_{\text{ax}} < H(A,X) \label{eq:FSPDUniformity}
\end{align}
by \cite[Theorem 1]{OSRBAmin}.

By \cite[Lemma 1]{OSRBAmin} we get reliability, i.e., the eavesdropper can recover $(a^n,x^n)$ from $(f_{\text{ax}},y_1^n,s_1^n)$, when 
\begin{align}
    \widetilde{R}_{\text{ax}} > H(A,X|Y_1,S_1). \label{eq:FSPDReliability}
\end{align}

Security follows when 
\begin{align}
    \overbar{R}_{\text{y}_1} = \overbar{R}_{\text{ax}} < H(Y_1|A,X,Y_2,S_2) \label{eq:FSPDSec1}
\end{align}
since $L_{\text{y}_1}$ becomes almost independent of $(A^n,X^n,Y_2,S_2)$ by \cite[Theorem 1]{OSRBAmin}. Similarly, by \cite[Theorem 1]{OSRBAmin} $F_{\text{ax}}$ and $W_{\text{ax}}$ are uniformly distributed and independent of $(Y_2^n,S_2^n)$ when
\begin{align}
    \widetilde{R}_{\text{ax}} + R_{\text{ax}} < H(A,X|Y_2,S_2).\label{eq:FSPDSec2}
\end{align}

Performing Fourier-Motzkin Elimination on \eqref{eq:FSPDUniformity}~-~\eqref{eq:FSPDSec2}, we have
\begin{align}
    &R_{\text{ax}} + \overbar{R}_{\text{ax}} < I(A,X;Y_1,S_1)\\
    &R_{\text{ax}} < \big[H(A,X|Y_2,S_2)- H(A,X|Y_1,S_1)\big]^+\\
    &\overbar{R}_{\text{ax}} < H(Y_1|Y_2,S_2,A,X).
\end{align}
Choosing $R=R_{\text{ax}} + \overbar{R}_{\text{ax}}$, we have that for any $\epsilon>0$ 
\begin{align}
    &R = \min\big\{I(A,X;Y_1,S_1),\nonumber\\
    &\qquad\qquad\quad \big(\big[H(A,X|Y_2,S_2)- H(A,X|Y_1,S_1)\big]^+\nonumber\\
    &\qquad\qquad\qquad   + H(Y_1|Y_2,S_2,A,X)\big)\big\}-\epsilon \label{eq:FSPDRConstraintCombined}
\end{align}
is achievable. Using the physical-degradation, we can simplify the rate condition  
\begin{align}
    &\big[H(A,X|Y_2,S_2)- H(A,X|Y_1,S_1)\big]^+ + H(Y_1|Y_2,S_2,A,X)\nonumber\\
    &\overset{(a)}{=} H(A,X|Y_2,S_2)- H(A,X|Y_1,S_1) + H(Y_1|Y_2,S_2,A,X)\nonumber\\
    & \overset{(b)}{=} H(A,X,Y_1|Y_2,S_2) - H(A,X|Y_1,S_1,Y_2,S_2)\nonumber\\
    & = H(Y_1|Y_2,S_2) + I(A,X;S_1|Y_1,Y_2,S_2)\nonumber\\
    & = H(Y_1,S_1|Y_2,S_2)-H(S_1|Y_1,Y_2,S_2,X,A) \label{eq:FSPDRConstraintP2}
\end{align}
where $(a)$ and $(b)$ follow since $(A,X)-(Y_1,S_1)-(Y_2,S_2)$ forms a Markov chain. The rate constraint in the theorem, \eqref{eq:FSachPD2} follows by combining \eqref{eq:FSPDRConstraintP2} with \eqref{eq:FSPDRConstraintCombined}.

The distortion constraints, optimality of per-letter deterministic estimators, derandomization of Protocol B, and the removal of the secret key and subsequent block-Markov coding scheme secrecy analysis follow from the same argument Appendix~\ref{app:thm1}.

\noindent\textbf{Converse:}
We assume that for some $\delta_n>0$, with $\delta_n\rightarrow0$ as $n\rightarrow\infty$, there exists an action encoder, decoder, and estimators such that the rate, reliability, and secrecy conditions \eqref{eq:FSrates_cons}-\eqref{eq:FSsecrecyleakage_cons} and the distortion constraints \eqref{eq:distortion_cons} are satisfied for $(R,D_1,D_2)$.

By Fano's inequality we obtain
\begin{align}
    H(M;Y_1^n,S_1^n)\leq n\epsilon_n \label{eq:FSFano}
\end{align}
where $n\epsilon_n=H_b(\delta_n)+\delta_n(R)n$, noting that $\epsilon_n\rightarrow0$ as $\delta_n\rightarrow0$.

We next outline the two bounds on $R$.
\begin{align}
    &nR\overset{(a)}{\leq} I(M;Y_1^n,S_1^n)\nonumber\\
    &=\sum\limits_{i=1}^n\big[H(Y_{1,i},S_{1,i}|Y_1^{i-1},S_1^{i-1})\nonumber\\
    &\qquad\qquad-H(Y_{1,i},S_{1,i}|Y_1^{i-1},S_1^{i-1},M)\big] + n\epsilon_n\nonumber\\
    &\leq\sum\limits_{i=1}^n\big[H(Y_{1,i},S_{1,i})\nonumber\\
    &\qquad\qquad-H(Y_{1,i},S_{1,i}|Y_1^{i-1},S_1^{i-1},M,X_i,A_i)\big] + n\epsilon_n\nonumber\\
    &\overset{(b)}{=}\sum\limits_{i=1}^nI(X_i,A_i;Y_{1,i},S_{1,i})+n\epsilon_n
\end{align}
where $(a)$ follows from Fano's inequality \eqref{eq:FSFano}, and $(b)$ is a result of the Markov chain 
\begin{align*}
    (Y_{1,i},S_{1,i})-(X_i,A_i)-(Y_1^{i-1},S_1^{i-1},M).
\end{align*}

We can also bound $R$ as
\begin{align}
    &nR \overset{(a)}{\leq} I(M;Y_1^n,S_1^n)+n\epsilon_n\nonumber\\
    & \leq I(M;Y_1^n,S_1^n,Y_2^n,S_2^n) +n\epsilon_n\nonumber\\
    & = H(Y_1^n,S_1^n|Y_2^n,S_2^n) + H(Y_2^n,S_2^n)- H(Y_2^n,S_2^n|M)\nonumber\\
    &\qquad-H(Y_1^n,S_1^n|M,Y_2^n,S_2^n)+n\epsilon_n\nonumber\\
    & = H(Y_1^n,S_1^n|Y_2^n,S_2^n) + I(M;Y_2^n,S_2^n)\nonumber\\
    &\qquad-H(Y_1^n,S_1^n|M,Y_2^n,S_2^n)+n\epsilon_n\nonumber\\
    & \overset{(b)}{\leq} H(Y_1^n,S_1^n|Y_2^n,S_2^n) + \delta_n\nonumber\\
    &\qquad-H(S_1^n|M,Y_1^n,Y_2^n,S_2^n)+n\epsilon_n\nonumber\\
    &\leq\sum\limits_{i=1}^n\big[H(Y_{1,i},S_{1,i}|Y_{2,i},S_{2,i}) +\frac{1}{n}\delta_n + \epsilon_n\nonumber\\
    &\qquad\qquad-H(S_{1,i}|Y_{1,i}^n,Y_{2,i}^n,S_{2,i}^n,M,S_1^{i-1},X_i,A_i)\big]\nonumber\\
    & \overset{(c)}{=}\sum\limits_{i=1}^n\big[H(Y_{1,i},S_{1,i}|Y_{2,i},S_{2,i})+\frac{1}{n}\delta_n + \epsilon_n\nonumber\\
    &\qquad\qquad -H(S_{1,i}|Y_{1,i},Y_{2,i},S_{2,i},X_i,A_i)\big] 
\end{align}
where $(a)$ follows by Fano's inequality \eqref{eq:FSFano}, $b$ follows from the secrecy constraint \eqref{eq:FSsecrecyleakage_cons} and $(c)$ follows by application of the Markov chain 
\begin{align*}
    S_{1,i}\!-\!(Y_{1,i},Y_{2,i},S_{2,i},X_i,A_i)\!-\!(Y_1^{n\setminus i},Y_2^{n\setminus i},S_2^{n\setminus i},M,S_1^{i-1}).
\end{align*}

Next, we introduce a time-sharing random variable $Q$ distributed uniformly on $[1:n]$ which is independent of all other random variables. This allows us to represent the bounds on $R$ as follows
\begin{align}
    &R \leq \frac{1}{n}\sum\limits_{i=1}^nI(X_i,A_i;Y_{1,i},S_{1,i}) + \epsilon_n\nonumber\\
    & = \frac{1}{n}\sum\limits_{i=1}^nI(X_Q,A_Q;Y_{1,Q},S_{1,Q}|Q=i) + \epsilon_n\nonumber\\
    & = I(X_Q,A_Q;Y_{1,Q},S_{1,Q}|Q) + \epsilon_n
\end{align}
and
\begin{align}
    & R \leq \frac{1}{n}\sum\limits_{i=1}^n\big[H(Y_{1,i},S_{1,i}|Y_{2,i},S_{2,i})+\frac{1}{n}\delta_n + \epsilon_n\nonumber\\
    &\qquad\qquad\quad -H(S_{1,i}|Y_{1,i},Y_{2,i},S_{2,i},X_i,A_i)\big] \nonumber\\
    &= \frac{1}{n}\sum\limits_{i=1}^n\big[H(Y_{1,Q},S_{1,Q}|Y_{2,Q},S_{2,Q},Q=i) +\frac{1}{n}\delta_n + \epsilon_n\nonumber\\
    &\qquad\qquad\quad -H(S_{1,Q}|Y_{1,Q},Y_{2,Q},S_{2,Q},X_Q,A_Q,Q=i)\big]\nonumber\\
    &= H(Y_{1,Q},S_{1,Q}|Y_{2,Q},S_{2,Q},Q)\nonumber\\
    &\quad -H(S_{1,Q}|Y_{1,Q},Y_{2,Q},S_{2,Q},X_Q,A_Q,Q) +\frac{1}{n}\delta_n + \epsilon_n\nonumber\\
    &\overset{(a)}{\leq} H(Y_{1,Q},S_{1,Q}|Y_{2,Q},S_{2,Q})\nonumber\\
    &\quad -H(S_{1,Q}|Y_{1,Q},Y_{2,Q},S_{2,Q},X_Q,A_Q) +\frac{1}{n}\delta_n + \epsilon_n
\end{align}
where $(a)$ follows from $Q-(A_Q,X_Q)-(Y_{1,Q},S_{1,Q},Y_{2,Q},S_{2,Q})$. Defining $X=X_Q$, $A=A_Q$, $Y_1=Y_{1,Q}$, $S_1=S_{1,Q}$, $Y_1=Y_{1,Q}$, $S_2=S_{2,Q}$, and $V=(V_Q,Q)$ so that $V-(X,A)-(Y_1,S_1,Y_2,S_2)$ forms a Markov chain and letting $\delta_n\rightarrow0$ gives
\begin{align}
    R & \leq \min\big\{I(X,A;Y_{1},S_{1}|Q),\nonumber\\
    &\qquad\qquad H(Y_{1},\!S_{1}|Y_{2},\!S_{2})\!-\!H(S_{1}|Y_{1},\!Y_{2},\!S_{2},\!X\!,\!A)\!\big\}.
\end{align}
Since we only need to preserve $I(X,A;Y_{1},S_{1}|Q)$, we can bound $|\mathcal{Q}|$ by $1$ by the support lemma \cite[Lemma 15.4]{CsiszarKornerbook2011}, resulting in the rate given in the theorem, \eqref{eq:FSachPD2}.

The distortion constraints follow as in Appendix~\ref{app:thm1}.
\end{IEEEproof}

\section{Proof of Theorem~\ref{theo:RPDfullsecrecy}}\label{app:thm4}
\begin{IEEEproof}
\textbf{Achievability:}
The construction and analysis of Protocols A and B follow as in Appendix~\ref{app:thm3} reach the constraint \eqref{eq:FSPDRConstraintCombined} with out the use of degradation, i.e.,
\begin{align}
    &R = \min\big\{I(A,X;Y_1,S_1),H(Y_1|Y_2,S_2,A,X) +\nonumber\\
    &\qquad\qquad\quad\big[\!H(A,\!X|Y_2,\!S_2)\!-\! H(\!A,\!X|Y_1,\!S_1)\big]^+\! \big\}\!-\!\epsilon \label{eq:FSRPDRConstraintCombined}
\end{align}
is achievable. Using reverse-physical-degradation, see \eqref{eq:rpdmc}, we can simplify the rate condition  
\begin{align}
    &H(Y_1|Y_2,S_2,A,X) + \big[H(A,X|Y_2,S_2)- H(A,X|Y_1,S_1)\big]^+\nonumber\\
    & \overset{(a)}{=} H(Y_1|Y_2,S_2,A,X)\nonumber\\
    & \overset{(b)}{=} H(Y_1|Y_2,S_2) \label{eq:FSRPDRConstraintP2}
\end{align}
where $(a)$ and $(b)$ follow since $(A,X)-(Y_2,S_2)-(Y_1,S_1)$ forms a Markov chain. The rate constraint in the theorem, \eqref{eq:FSachRPD2} follows by combining \eqref{eq:FSRPDRConstraintP2} with \eqref{eq:FSRPDRConstraintCombined}.

The distortion constraints and block-Markov coding scheme analysis follow as in Appendix~\ref{app:thm3}.

\noindent\textbf{Converse:} The converse of Theorem~\ref{theo:PDfullsecrecy} does not use degradation. We apply the reverse-physical-degradation to \eqref{eq:FSachPD2} by performing the following simplification
\begin{align}
    &H(Y_{1},S_{1}|Y_{2},S_{2})-H(S_{1}|Y_{1},Y_{2},S_{2},X,A)\nonumber\\
    & \overset{(a)}{=} H(Y_{1},S_{1}|Y_{2},S_{2}) - H(S_{1}|Y_{1},Y_{2},S_{2})\nonumber\\
    & = H(Y_1|Y_2,S_2)\label{eq:FSRPDRConstraintSimplification}
\end{align}
where $(a)$ follows since $(A,X)-(Y_2,S_2)-(Y_1,S_1)$. Combining \eqref{eq:FSachPD2} with \eqref{eq:FSRPDRConstraintSimplification} gives the rate condition in the theorem statement, \eqref{eq:FSachRPD2}.

\end{IEEEproof}

\section{Proof of Lemma~\ref{lem:BinaryExample}}\label{app:lem1}
\begin{IEEEproof}
The lemma follows from evaluating the strong-secrecy distortion region $R_{Act}$ defined in Theorem~\ref{theo:PDfullsecrecy}. We have
\begin{align}
    &H(Y_1,S_1|Y_2,S_2)-H(S_1|Y_1,Y_2,S_2,X,A)\nonumber\\
    &= H(S_1|Y_2,S_2)+H(Y_1|S_1,Y_2,S_2)-H(S_1|Y_1,S_2,X,A)
\end{align}
which we evaluate term by term.

For the first term we have
\begin{align}
    H(S_1|Y_2,S_2) & \overset{(a)}{=} H(S_1|Y_2,S_2=0)p_{S_2}(0)\nonumber
        \end{align}
    \begin{align}
    & = H(S_1|Y_2=1,S_2=0)p_{Y_2S_2}(1,0) \nonumber\\
    &\qquad+ H(S_1|Y_2=0,S_2=0)p_{Y_2S_2}(0,0)\nonumber\\
    & \overset{(b)}{=} H(S_1|Y_2=0,S_2=0)\nonumber\\
    & \overset{(c)}{=} H(S_1|S_2=0)\nonumber\\
    & = H_b(p_{S_1|S_2}(0,0))\nonumber\\
    & = H_b\bigg(\frac{1-\lambda}{1-\alpha\lambda}\bigg) \label{eq:lemR1p1}
\end{align}
where $(a)$ follows since $H(S_1|Y_2,S_2=1)=0$ because $S_1=1$ if $S_2=1$, $(b)$ follows since $p_{Y_2S_2}(1,0)=0$ and $p_{Y_2S_2}(0,0)=1$, and $(c)$ follows since $Y_2=0$ if $S_2=0$.

Evaluating the second term gives us
\begin{align}
    &H(Y_1|S_1,Y_2,S_2) \nonumber\\
    & \overset{(a)}{=} H(Y_1|S_1=1,Y_2,S_2)p_{S_1}(1)\nonumber\\
    & \overset{(b)}{=} H(Y_1|S_1=1,Y_2,S_2=0)p_{S_1S_2}(1,0)\nonumber\\
    & = H(Y_1|S_1=1,Y_2=0,S_2=0)p_{S_1Y_2S_2}(1,0,0)\nonumber\\
    &\qquad +H(Y_1|S_1=1,Y_2=1,S_2=0)p_{S_1Y_2S_2}(1,1,0)\nonumber\\
    & \overset{(c)}{=} H(Y_1|S_1=1,S_2=0)p_{S_1S_2}(1,0)\nonumber\\
    & \overset{(d)}{=} H(X|S_1=1)p_{S_1S_2}(1,0)\nonumber\\
    & = H_b(p_{X|S_1}(1|1))p_{S_1S_2}(1,0) \nonumber\\
    & = H_b\bigg(\frac{p(1-q\ast\lambda)}{1-\lambda\ast p\ast q}\bigg)(1-\alpha)(1-\lambda\ast p\ast q) \label{eq:lemR1p2}
\end{align}
where $(a)$ follows since $H(Y_1|S_1=0,Y_2,S_2)=0$ because $Y_1=0$ when $S_1=0$, $(b)$ follows since $H(Y_1|S_1=1,Y_2,S_2=1)=0$ because $Y_1=Y_2$ if $S_1=S_2=1$, $(c)$ follows since $Y_2$ must be $0$ if $S_2=0$, and $(d)$ follows since $(X,A)-S_1-S_2$ forms a Markov chain and $Y_1 = S_1 \cdot X = X$.

The third term is
\begin{align}
    &H(S_1|Y_1,Y_2,S_2,X,A)\nonumber\\
    & \overset{(a)}{=} H(S_1|Y_1,Y_2,S_2=0,X,A)p_{S_2}(0)\nonumber\\
    & \overset{(b)}{=} H(S_1|Y_1=0,Y_2,S_2=0,X,A)p_{Y_1S_2}(0,0)\nonumber\\
    & \overset{(c)}{=} H(S_1|Y_1=0,Y_2=0,S_2=0,X,A)p_{Y_1Y_2S_2}(0,0,0)\nonumber\\
    & \overset{(d)}{=} H(S_1|Y_1\!=\!0,Y_2\!=\!0,S_2\!=\!0,X\!=\!0,A)p_{XY_1Y_2S_2}(0,0,0,0)\nonumber\\
    & \overset{(e)}{=} H(S_1|X\!=\!0,A,S_2\!=\!0)p_{XS_2}(0,0)\nonumber\\
    & = H(S_1|X=0,A=0,S_2=0)p_{XAS_2}(0,0,0) \nonumber\\
    &\qquad+ H(S_1|X=0,A=1,S_2=0)p_{XAS_2}(0,1,0)\nonumber\\
    & = H_b(p_{S_1|XAS_2}(0|0,0,0))p_{XAS_2}(0,0,0)\nonumber\\
    &\qquad+ H_b(p_{S_1|XAS_2}(0|0,1,0))p_{XAS_2}(0,1,0)\nonumber\\
    & = H_b\bigg(\frac{\lambda}{1-\alpha+\lambda\alpha}\bigg)(1-p)(1-q)(1-\alpha+\alpha\lambda)\nonumber\\
    &\qquad + H_b\bigg(\frac{1-\lambda}{1-\lambda\alpha}\bigg)(1-p)q(1-\lambda\alpha) \label{eq:lemR1p3}
\end{align}
where $(a)$ follows since $H(S_1|Y_1,Y_2,S_2=1,X,A)=0$ because $S_1=1$ if $S_2=1$, $(b)$ follows since we have 
\begin{align}
H(S_1|Y_1=1,Y_2,S_2=0,X,A)=0
\end{align}
because $S_1=1$ if $Y_1=1$, $(c)$ follows since $p_{Y_1Y_2S_2}(0,1,0)=0$ because $Y_2$ cannot be $1$ if $S_2=0$, $(d)$ follows since $H(S_1|Y_1=0,Y_2=0,S_2=0,X=1,A)=0$ because $S_1=0$ if $Y_1=0$ and $X=1$, and $(e)$ follows since $Y_1$ and $Y_2$ must be zero if $X=0$.

Combining \eqref{eq:lemR1p1}-\eqref{eq:lemR1p3} gives the first term in the minimization in \eqref{eq:lemRcondition}.

We now calculate the second term in \eqref{eq:lemRcondition}, $I(X,A;Y_1,S_1) = H(X,A) - H(X,A|Y_1,S_1)$ in two parts. The first part is given by
\begin{align}
    &H(X,A) \nonumber\\
    &= -(1\!-\!p)(1\!-\!q)\log\big((1\!-\!p)(1\!-\!q)\big) - pq\log\big(pq\big)\nonumber\\
    & \qquad -p(1\!-\!q)\log\big(p(1\!-\!q)\big) - (1\!-\!p)q\log\big((1\!-\!p)q\big).\label{eq:lemR2p1}
\end{align}

The second part is 
\begin{align}
    &H(X,A|Y_1,S_1) \nonumber\\ 
    &\overset{(a)}{=} H(X,A|Y_1,S_1=0)p_{S_1}(0) + H(A|X,S_1=1)p_{S_1}(1) \nonumber\\
    &\overset{(b)}{=} H(X,A|S_1=0)p_{S_1}(0) + H(A|X,S_1=1)p_{S_1}(1) \nonumber\\
    & = H(X,A|S_1\!=\!0)p_{S_1}(0) \!+\! H\!(A|X\!=\!0,S_1\!=\!1)p_{XS_1}(0,1)\nonumber\\
    &\qquad + H\!(A|X\!=\!1,S_1\!=\!1)p_{XS_1}(1,1)\nonumber\\
    & = H(X,A|S_1\!=\!0)p_{S_1}(0) \!+\! H_b\big(p_{A|XS_1}(1|0,\!1)\big)p_{XS_1}(0,\!1)\nonumber\\
    &\qquad + H_b\big(p_{A|XS_1}(0|1,\!1)\big)p_{XS_1}(1,\!1)\nonumber\\
    & = -(1\!-\!p)(1\!-\!q)\lambda\log\frac{(1\!-\!p)(1\!-\!q)\lambda}{\lambda\ast p\ast q} \nonumber\\
    &\qquad-(1\!-\!p)q(1\!-\!\lambda)\log\frac{(1\!-\!p)q(1\!-\!\lambda)}{\lambda\ast p\ast q}\nonumber\\
    &\qquad-pq\lambda\log\frac{pq\lambda}{\lambda\ast p\ast q} -p(1\!-\!q)(1\!-\!\lambda)\log\frac{p(1\!-\!q)(1\!-\!\lambda)}{\lambda\ast p\ast q}\nonumber\\
    &\qquad +\! H_b\!\bigg(\!\frac{q\lambda}{1\!-\!q\ast\lambda}\!\bigg)\!(1\!-\!p)(1\!-\!q\ast\lambda) \!+\! H_b\!\bigg(\!\frac{q(1\!-\!\lambda)}{q\ast\lambda}\!\bigg)\!p(q\ast\lambda)\label{eq:lemR2p2}
\end{align}
where $(a)$ follows since $Y_1=X$ when $S_1=1$, and $H(X,A|X,S_1=1)=H(A|X,S_1=1)$ and $(b)$ follows since $Y_1=X\cdot S_1=X\cdot 0=0$ is deterministic and thus is independent of $(X,A)$. Combining \eqref{eq:lemR2p1} and \eqref{eq:lemR2p2} gives the second term in the minimization in \eqref{eq:lemRcondition}.

Now we calculate the distortion constraints. We will use estimators of the form $\Est_j(x,a,y_j)$. Consider the case where $X=1$. Then we have $Y_1=S_1\cdot X=S_1$. The estimate $\Est_1(1,a,y_1)=y_1$ will always be correct. Consider $Y_2=S_2\cdot Y_1 = S_2\cdot S_1\cdot X=S_2\cdot S_1$, then choosing $\Est_2(1,a,y_2)=y_2$ is also error free because $S_2$ cannot be $1$ if $S_1=0$.

Now we consider the case where $X=0$. When estimating $Y_1$, we get $\Est_1(0,0,y_1)=\mathbbm{1}\{\lambda < 0.5\}$ and $\Est_1(0,1,y_1)\!=\!\mathbbm{1}\{\lambda\! \geq\! 0.5\}$, equivalent to $\Est_1(0,a,y_1)\!=\!a\!-\!\mathbbm{1}\{\lambda\! <\!0.5\}$. When estimating $Y_2$, the optimal estimator is equal to $\Est_2(0,0,y_2)=\mathbbm{1}\{\alpha-\alpha\lambda > 0.5\}$ and $\Est_1(0,1,y_1)=\mathbbm{1}\{\alpha\lambda > 0.5\}$. Combining these gives
\begin{align}
    \Est_1(x,a,y_1) = \begin{cases}
y_1  & \mbox{if }x=0\\
a-\mathbbm{1}\{\lambda <.5\} & \mbox{if }x=1
\end{cases}
\end{align}
and
\begin{align}
\Est_1(x,a,y_1) = \begin{cases}
    y_1  & \mbox{if }x=0\\
    \mathbbm{1}\{\alpha-\alpha\lambda >.5\} & \mbox{if }x=1,a=0\\
    \mathbbm{1}\{\alpha\lambda >.5\} & \mbox{if }x=1,a=1.
\end{cases}
\end{align}

Using the Hamming distortion metric, the expected distortion for $S_1$ is \eqref{eq:lemD1} and the expected distortion for $S_2$ is equal to \eqref{eq:lemD2}.
\end{IEEEproof}

\end{document}